\pdfoutput=1
\documentclass[twocolumn]{aastex631}

\usepackage{comment}
\usepackage{bm}
\usepackage{multirow}
\usepackage{appendix}

\revised{December 28, 2021}
\accepted{December 28, 2021}
\submitjournal{Astronomical Journal}

\shorttitle{OGLE-2014-BLG-0319}
\shortauthors{Miyazaki et al.}

\begin{document}

\title{OGLE-2014-BLG-0319: A Sub-Jupiter-Mass Planetary Event Encountered Degeneracy with Different Mass Ratios and Lens-Source Relative Proper Motions}

\author[0000-0001-9818-1513]{Shota Miyazaki}
\altaffiliation{MOA collaboration}
\affiliation{Department of Earth and Space Science, Graduate School of Science, Osaka University, 1-1 Machikaneyama, Toyonaka, Osaka 560-0043, Japan}
\author[0000-0001-9818-1513]{Daisuke Suzuki}
\altaffiliation{MOA collaboration}
\affiliation{Department of Earth and Space Science, Graduate School of Science, Osaka University, 1-1 Machikaneyama, Toyonaka, Osaka 560-0043, Japan}
\author{Andrzej Udalski}
\altaffiliation{OGLE collaboration}
\affiliation{Astronomical Observatory, University of Warsaw, AI. Ujazdowskie 4, 00-478 Warszawa, Poland}
\author{Naoki Koshimoto}
\altaffiliation{MOA collaboration}
\affiliation{Code 667, NASA Goddard Space Flight Center, Greenbelt, MD 20771, USA}
\affiliation{Department of Astronomy, University of Maryland, College Park, MD 20742, USA}
\author[0000-0001-8043-8413]{David P. Bennett}
\altaffiliation{MOA collaboration}
\affiliation{Code 667, NASA Goddard Space Flight Center, Greenbelt, MD 20771, USA}
\affiliation{Department of Astronomy, University of Maryland, College Park, MD 20742, USA}
\author[0000-0001-5069-319X]{Nicholas Rattenbury}
\altaffiliation{MOA collaboration}
\affiliation{Department of Physics, University of Auckland, Private Bag 92019, Auckland, New Zealand}
\author{Takahiro Sumi}
\altaffiliation{MOA collaboration}
\affiliation{Department of Earth and Space Science, Graduate School of Science, Osaka University, 1-1 Machikaneyama, Toyonaka, Osaka 560-0043, Japan}
\collaboration{100}{(Leading authors)}

\author{Fumio Abe}
\affiliation{Institute for Space-Earth Environmental Research, Nagoya University, Nagoya 464-8601, Japan}
\author{Richard K. Barry}
\affiliation{Code 667, NASA Goddard Space Flight Center, Greenbelt, MD 20771, USA}
\author{Aparna Bhattacharya}
\affiliation{Department of Physics, University of Notre Dame, Notre Dame, IN 46556, USA} 
\affiliation{Code 667, NASA Goddard Space Flight Center, Greenbelt, MD 20771, USA}
\author{Ian A. Bond}
\affiliation{Institute of Information and Mathematical Sciences, Massey University, Private Bag 102-904, North Shore Mail Centre, Auckland, New Zealand}
\author{Akihiko Fukui}
\affiliation{Department of Earth and Planetary Science, Graduate School of Science, The University of Tokyo, 7-3-1 Hongo, Bunkyo-ku, Tokyo 113-0033, Japan}
\affiliation{Instituto de Astrof\'isica de Canarias, V\'ia L\'actea s/n, E-38205 La Laguna, Tenerife, Spain}
\author{Hirosane Fujii}
\affiliation{Department of Earth and Space Science, Graduate School of Science, Osaka University, 1-1 Machikaneyama, Toyonaka, Osaka 560-0043, Japan}
\author{Yuki Hirao}
\affiliation{Department of Earth and Space Science, Graduate School of Science, Osaka University, 1-1 Machikaneyama, Toyonaka, Osaka 560-0043, Japan}
\author[0000-0003-2267-1246]{Stela Ishitani~Silva}
\affiliation{Code 667, NASA Goddard Space Flight Center, Greenbelt, MD 20771, USA}
\affiliation{Department of Physics, The Catholic University of America, Washington, DC 20064, USA}
\author{Yoshitaka Itow}
\affiliation{Institute for Space-Earth Environmental Research, Nagoya University, Nagoya, 464-8601, Japan}
\author{Rintaro Kirikawa}
\affiliation{Department of Earth and Space Science, Graduate School of Science, Osaka University, 1-1 Machikaneyama, Toyonaka, Osaka 560-0043, Japan}
\author{Iona Kondo}
\affiliation{Department of Earth and Space Science, Graduate School of Science, Osaka University, 1-1 Machikaneyama, Toyonaka, Osaka 560-0043, Japan}
\author{Brandon Munford}
\affiliation{Department of Physics, University of Auckland, Private Bag 92019, Auckland, New Zealand}
\author{Yutaka Matsubara}
\affiliation{Institute for Space-Earth Environmental Research, Nagoya University, Nagoya, 464-8601, Japan}
\author{Sho Matsumoto}
\affiliation{Department of Earth and Space Science, Graduate School of Science, Osaka University, 1-1 Machikaneyama, Toyonaka, Osaka 560-0043, Japan}
\author{Yasushi Muraki}
\affiliation{Institute for Space-Earth Environmental Research, Nagoya University, Nagoya, 464-8601, Japan}
\author{Arisa Okamura}
\affiliation{Department of Earth and Space Science, Graduate School of Science, Osaka University, 1-1 Machikaneyama, Toyonaka, Osaka 560-0043, Japan}
\author{Greg Olmschenk}
\affiliation{Code 667, NASA Goddard Space Flight Center, Greenbelt, MD 20771, USA}
\affiliation{Universities Space Research Association, Columbia, MD 21046, USA}
\author[0000-0003-2388-4534]{Cl\'{e}ment Ranc}
\affiliation{Sorbonne Universit\'e, CNRS, UMR 7095, Institut d’Astrophysique de Paris, 98 bis bd Arago, 75014 Paris, France}
\author{Yuki K. Satoh}
\affiliation{Department of Earth and Space Science, Graduate School of Science, Osaka University, 1-1 Machikaneyama, Toyonaka, Osaka 560-0043, Japan}
\author{Taiga Toda}
\affiliation{Department of Earth and Space Science, Graduate School of Science, Osaka University, 1-1 Machikaneyama, Toyonaka, Osaka 560-0043, Japan}
\author{Paul J. Tristram}
\affiliation{University of Canterbury Mt. John Observatory, P.O. Box 56, Lake Tekapo 8770, New Zealand}
\author{Hibiki Yama}
\affiliation{Department of Earth and Space Science, Graduate School of Science, Osaka University, 1-1 Machikaneyama, Toyonaka, Osaka 560-0043, Japan}
\author{Atsunori Yonehara}
\affiliation{Department of Physics, Faculty of Science, Kyoto Sangyo University, Kyoto 603-8555, Japan}
\collaboration{100}{(MOA collaboration)}

\author{Radek Poleski}
\affiliation{Astronomical Observatory, University of Warsaw, AI. Ujazdowskie 4, 00-478 Warszawa, Poland}
\author{Przemek Mr\'{o}z}
\affiliation{Livision of physics, Mathematics, and Astronomy, California institute of Technology, Pasadena, CA 91125, USA}
\author{Jan Skowron}
\affiliation{Astronomical Observatory, University of Warsaw, AI. Ujazdowskie 4, 00-478 Warszawa, Poland}
\author{Michal K. Szyma\'{n}ski}
\affiliation{Astronomical Observatory, University of Warsaw, AI. Ujazdowskie 4, 00-478 Warszawa, Poland}
\author{Igor Soszy\'{n}ski}
\affiliation{Astronomical Observatory, University of Warsaw, AI. Ujazdowskie 4, 00-478 Warszawa, Poland}
\author{Pawel Pietrukowicz}
\affiliation{Astronomical Observatory, University of Warsaw, AI. Ujazdowskie 4, 00-478 Warszawa, Poland}
\author{Syzmon Koz\l owski}
\affiliation{Astronomical Observatory, University of Warsaw, AI. Ujazdowskie 4, 00-478 Warszawa, Poland}
\author{Krzysztof Ulaczyk}
\affiliation{Astronomical Observatory, University of Warsaw, AI. Ujazdowskie 4, 00-478 Warszawa, Poland}
\affiliation{Department of Physics, University of Warwick, Gibbet Hill Road, Coventry, CV4 7AL, UK}
\author{{\L}ukasz Wyrzykowski}
\affiliation{Astronomical Observatory, University of Warsaw, AI. Ujazdowskie 4, 00-478 Warszawa, Poland}
\collaboration{100}{(OGLE collaboration)}

\begin{abstract}
We report the discovery of a sub-Jovian-mass planet, OGLE-2014-BLG-0319Lb.
The characteristics of this planet will be added into a future extended statistical analysis of the Microlensing Observations in Astrophysics (MOA) collaboration.
The planetary anomaly of the light curve is characterized by MOA and OGLE survey observations and results in three degenerate models with different planetary mass-ratios of $q=(10.3,6.6,4.5)\times10^{-4}$, respectively.
We find that the last two models require unreasonably small lens-source relative proper motions of $\mu_{\rm rel}\sim1\;{\rm mas/yr}$.
Considering Galactic prior probabilities, we rule out these two models from the final result.
We conduct a Bayesian analysis to estimate physical properties of the lens system using a Galactic model and find that the lens system is composed of a $0.49^{+0.35}_{-0.27}\;M_{\rm Jup}$ sub-Jovian planet orbiting a $0.47^{+0.33}_{-0.25}\;M_{\odot}$ M-dwarf near the Galactic bulge.
This analysis demonstrates that Galactic priors are useful to resolve this type of model degeneracy. 
This is important for estimating the mass ratio function statistically.
However, this method would be unlikely successful in shorter timescale events, which are mostly due to low-mass objects, like brown dwarfs or free-floating planets.
Therefore, careful treatment is needed for estimating the mass ratio function of the companions around such low-mass hosts which only the microlensing can probe.

\end{abstract}

\section{Introduction}
To date, more than 4000 exoplanets have been discovered \citep{Akeson+2013}, revealing the universality and the diversity of planetary systems.
Most of the known planets were discovered by the transit \citep{Charbonneau+2000} and the radial-velocity \citep[RV;][]{Butler+2006} methods that are relatively sensitive to planets massive and close to their host stars, and thus the distribution of close-orbit exoplanets within $\sim1\;{\rm au}$ have been revealed in detail by these methods \citep[e.g.,][]{Marcy+2005,Cumming+2008,Howard+2012,Fressin+2013}.
On the other hand, gravitational microlensing has a unique sensitivity to wide-orbit planets down to an Earth mass \citep{Gould+1992,Bennett+1996} beyond the snow line where planet formation is considered to be the most efficient in the core accretion model \citep{Lissauer1993}.
This sensitivity is complementary to that of other detection methods.

The most recent statistical analysis using the largest sample of microlensing planets was conducted by \citet{Suzuki+2016}, who studied six years of survey data from the second phase of the Microlensing Observations in Astrophysics \citep[MOA-II;][]{Bond+2001,Sumi+2003} collaboration, including 23 planets discovered from 1474 microlensing events in 2007-2012. 
They find that the planet-frequency function in the power law describing planet/host mass-ratio $q$ has a break around $q\sim2\times10^{-4}$, which implies Neptune-mass-ratio planets are the most abundant type of planet beyond the snow line.
Moreover, \citet{Suzuki+2018} compared the mass-ratio function with planet population synthesis models \citep{Ida+2004} and found a discrepancy over the range $10^{-3}<q<4\times10^{-3}$ of a factor of $\sim10$, suggesting there is no desert of sub-Saturn-mass planets beyond the snow line.
This is contrary to the prediction from the runaway accretion process in the core accretion model \citep[e.g.,][]{Pollack+1996}.
This result is also recently confirmed by the CORALIE/HARPS sample of planets found by the RV method \citep{Bennett+2021}.
These comparisons between observations and theories can provide opportunities to quantitatively diagnose sources of problems in calculation or theory \citep{Suzuki+2018}.
The MOA collaboration is planning to present new results of a statistical analysis using the extended sample of \citet{Suzuki+2016} including $\sim50$ planets detected by the MOA-II survey (D. Suzuki et al. in prep.).
OGLE-2014-BLG-0319Lb which is detailed in this paper will be entered into this statistical sample.

In general, planetary signals in microlensing events appear as short-lived anomalous deviations from typical single-lens light curves \citep{Paczynski1986,Mao+1991} and sometimes produce a degeneracy problem where several model interpretations are possible for an anomaly.
The origins of the degeneracy in microlensing events are summarized in detail in \citet{Han+2018}.
Identifying all possible models and degeneracies for each planetary event is important for the completeness of a statistical microlensing analysis because systematic analysis with automated parameter searches can sometimes miss the best-fit models for each events \citep[e.g., OGLE-2013-BLG-0911;][]{Shvartzvald+2016,Miyazaki+2020}.
Furthermore, investigating the origin of the degeneracy for each event would be important because this adds to the growing body of literature on this topic and thus helps avoid missing models in future analyses. 

In this paper, we analyze a microlensing event OGLE-2014-BLG-0319 that presents three planetary interpretations of different mass ratios of $q=(10.3,6.6,4.5)\times10^{-4}$ and investigate the origin of the model degeneracy.
The structure of this paper is as follows.
In Section \ref{sec:obs}, we describe the observation and the data sets of the event.
We present our light curve analysis and discuss the degeneracy in Section \ref{sec:model}.
We estimate physical properties of the lens and source systems in Section \ref{sec:physical} and finally summarize and discuss the result in Section \ref{sec:discussion}.

\section{Observations \& Data Reductions \label{sec:obs}}
\begin{figure*}
    \centering
    \includegraphics[scale=0.87,angle=270]{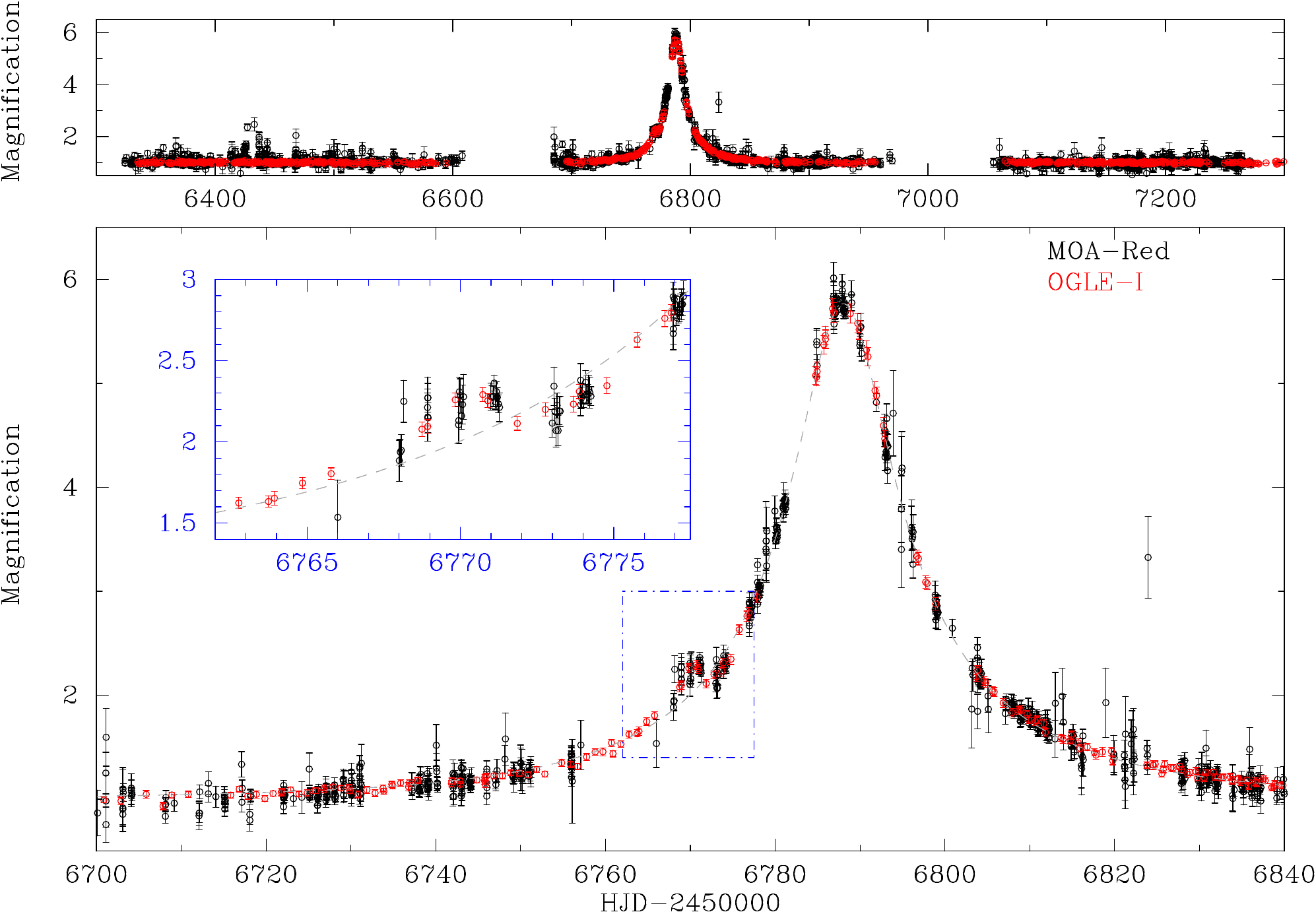}
    \caption{
    Light curve of OGLE-2014-BLG-0319 from the OGLE (red points) and MOA (black points) data.
    The gray dashed lines in panels represent the best-fit single-lens model.
    The blue panel shows a zoom-in view corresponding to the area surrounded by a blue chain line.
    }
    \label{fig:lc}
\end{figure*}

The microlensing event OGLE-2014-BLG-0319 occurred at $(\alpha,\delta)_{\rm J2000}=$(17:47:50.68, $-$33:56:06.16) which corresponds to $(l,b)=(-4.0306,-2.9971)$ in Galactic coordinates.
The fourth phase of the Optical Gravitational Lensing Experiment \citep[OGLE-IV;][]{Udalski+2015A} collaboration firstly discovered and reported the event on 2014 March 21 (about 50 days before the peak) through their Early Warning System \citep{Udalski2003}.
The OGLE field (BLG603) in which the event occurred was observed at a cadence of once per night by their 1.3m Warsaw telescope at Las Campanas Observatory in Chile using the OGLE-IV camera which has a 1.4 deg$^2$ field of view (FOV).
OGLE uses the standard Kron-Cousins \textit{I}-band filter in their regular surveys.
They also use the \textit{V}-band filter in their occasional observations in order to mainly measure source star color.
For this event, they did not conduct any $V$-band observations. 

Independently, the second phase of Microlensing Observations in Astrophysics (MOA-II) collaboration identified and reported the event as MOA-2014-BLG-171.
The MOA field (gb1) in which the event occurred was observed at a cadence of once per 45 minutes by their 1.8m MOA-II telescope at Mt. John Observatory in New Zealand using the MOA-cam3 camera which has a 2.2 deg$^{2}$ FOV \citep{Sako+2008}.
MOA uses in their regular survey observations a custom band-pass filter, named MOA-Red, which is similar to the sum of the standard Kron-Cousins $I$- and $R$-band filters. 
They also occasionally conduct $V$-band observations, but they did not observe in the $V$-band when the source star was magnified.

The data reductions for the OGLE and MOA data were conducted using their photometry pipelines \citep{Wozniak2000, Bond+2001} which are optimized in their implementations of the different image analysis \citep[DIA;][]{Alard+1998} method.
The OGLE photometry is calibrated to the standard Cousins $I$-band \citep{Szymanski+2011}.
In the MOA data, we found significant systematic variations in the light curve that are likely to be correlated with the seeing and airmass values.
We measured and corrected the systematic trends by running a detrending code \citep{Bond+2017}, which yielded a $\chi^2$ improvement of $\Delta\chi^2\sim0.3$ per data point in the baseline and thus significantly reduced the systematic trends in the MOA light curve.
We ran several of our detrending codes as used in \citet{Koshimoto+2017} and confirmed that these corrections are consistent with each other.
We also confirmed that this detrending procedure hardly affects the best-fit parameters for this event.  

In general, for crowded stellar regions toward the Galactic bulge, the error bars derived by each photometry pipeline are underestimated (or overestimated), which can lead to incorrect conclusions and uncertainties on physical parameters.
In order to correct this, we renormalize the error bars by $\sigma^{\prime}_{i}=k\sqrt{\sigma^{2}_{i}+e^{2}_{\rm min}}$, where $\sigma_{i}$ and $\sigma^{\prime}_{i}$ are the original and corrected error bars in magnitude respectively, and $k$ and $e_{\rm min}$ are renormalization coefficients \citep{Bennett+2008, Yee+2012}.
We select $k$ and $e_{\rm min}$ to satisfy $\chi^{2}$/dof=1 for each data set and make the cumulative $\chi^{2}$ distribution sorted by the source magnification as uniform a cumulative distribution as possible.
We first find a preliminary best-fit model using the original light curves.
Next, we renormalize the error bars applying $k$ and $e_{\rm min}$ values to satisfy the conditions referred above.
Finally, we find the final best-fit model fitting all the renormalized light curves.
We note that the best-fit model parameters and the final results are not sensitive to moderate changes of the renormalization factors \citep{Ranc+2019}. 
The data sets and the renormalization coefficients used in our analysis are summarized in Table \ref{tab:data}.

\begin{deluxetable}{ccccccc}
\tablecaption{Data Sets for OGLE-2014-BLG-0319\label{tab:data}}
\centering
\tablehead{
Label & Telescope & Passband & $N_{\rm data}$\tablenotemark{$^a$} & $k$\tablenotemark{$^b$} & $\sigma_{\rm min}$\tablenotemark{$^b$}
}
\startdata
MOA  & MOA-II 1.8m & MOA-Red   & 3147  & 1.891 & 0.003 \\
OGLE & Warsaw 1.3m & $I$       & 724   & 1.306 & 0.013 
\enddata
\tablenotetext{a}{The number of the data points.}
\tablenotetext{b}{The coefficients for the renormalization. See text.}
\end{deluxetable}

Figure \ref{fig:lc} shows the light curve from the OGLE and MOA data for OGLE-2014-BLG-0319.
Prior to the peak of the event, a significant deviation of the light curve from a standard single-lens single-source \citep[1L1S;][]{Paczynski1986} model was observed around ${\rm HJD}^\prime=6772$\footnote{${\rm HJD}^\prime={\rm HJD}-2450000$.}.
The anomaly was not noticed in real-time and was first announced on 2014 April 30 $({\rm HJD}^\prime=6778)$ when the deviation almost finished so that any follow-up observations were not conducted during the anomaly.
However, the deviation lasted about 10 days and both the OGLE and MOA regular observations could measure the deviation.
After that, some modelers in survey groups immediately performed binary-lens modeling for this event using the light curves produced by their real-time photometry pipelines which returns relatively rough DIA photometry\footnote{MOA, http://www.massey.ac.nz/~iabond/moa/alerts/}$^,$\footnote{OGLE, http://ogle.astrouw.edu.pl/ogle4/ews/ews.html}.
D. P. Bennett, one of the modelers, suggested in their private communication on 2014 May 17 that the anomaly could be explained with planetary-mass-ratio models although there is a potential model degeneracy, i.e. several competing models exist.

\section{Light-curve Modeling\label{sec:model}}

\begin{figure*}[t]
    \centering
    \begin{minipage}{0.45\hsize}
        \includegraphics[scale=0.5]{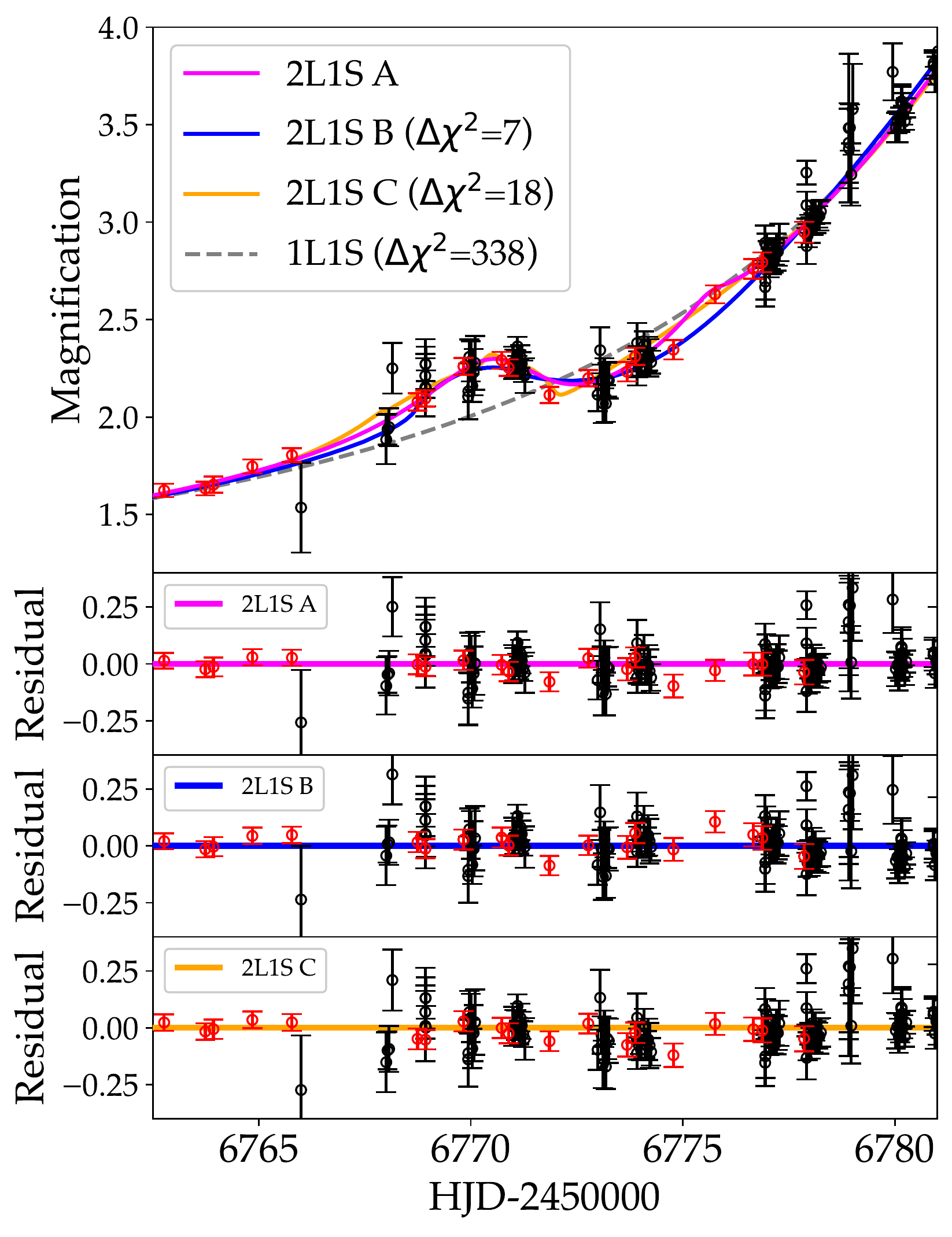}
    \end{minipage}
    \begin{minipage}{0.535\hsize}
        \includegraphics[scale=0.52]{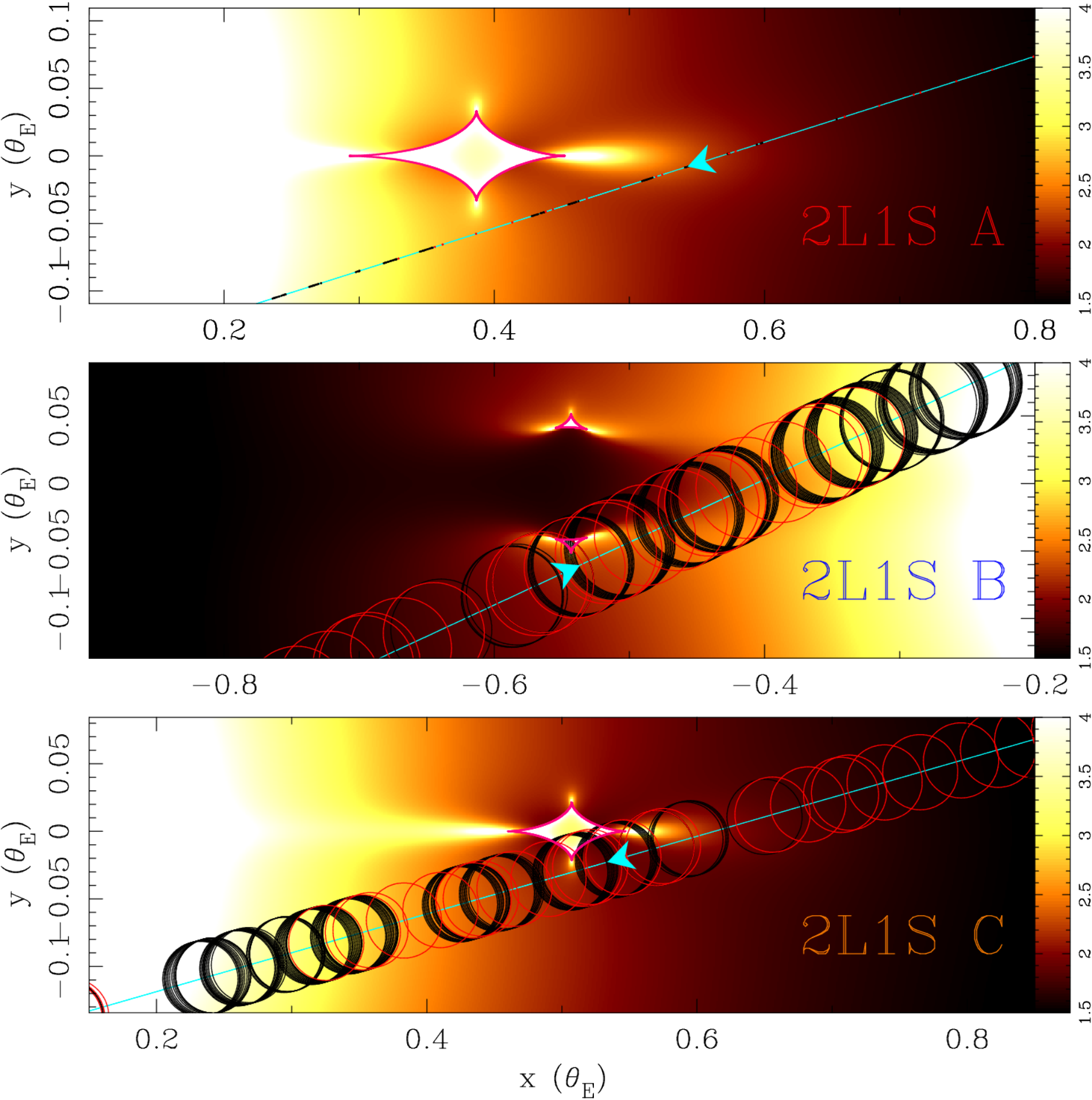}
    \end{minipage}
    \caption{
    Comparison of the three local solutions, models A, B, and C.
    ({\bf left panel})
    The left panel shows a detailed view of the light curves when the anomaly occurred.
    Each 2L1S model curve is presented by colored solid lines and the 1L1S model curve is shown as a dashed gray line.
    The black and red circles with error bars are the light curves of MOA and OGLE, respectively.
    ({\bf right panel})
    The right three panels show the geometries of planetary caustics (magenta solid lines) relative to the source trajectories (cyan solid lines) for each model.
    The cyan arrows indicate the directions in which the sources is moving. 
    The black and red circles represent the source positions when the MOA and OGLE measurements were conducted.
    The sizes of the circles are the best-fit source size of each model.
    The magnification maps are represented as color maps in each panel.
    }
    \label{fig:models}
\end{figure*}
As shown in Figure \ref{fig:lc}, the light curve of OGLE-2014-BLG-0319 shows a significant deviation from a \citet{Paczynski1986} single-lens single-source (1L1S) model.
Such a deviation could be produced by a single-lens binary-source (1L2S) model \citep{Gaudi1998,Shin+2019}.
However, we find that this 1L2S scenario is not preferred by $\Delta\chi^{2}>170$ comparing with the best-fit binary-lens single-source (2L1S) model\footnote{ The $\chi^2$ differences between the 2L1S and 1L1S models are $\sim 340$ for all the data and $\sim 220$ only for the MOA data.}.
The anomaly comprises of the bump and dip parts deviating from the 1L1S model, and is well-separated from the peak of the light curve, which implies the anomaly is likely due to the source crossing over or near planetary caustics.

Here we present an exploration of the best-fit model to explain the nature of OGLE-2014-BLG-0319.
Our light-curve modeling is done by our implementations of the image centered ray-shooting method \citep{Bennett+1996,Bennett2010} and the Markov Chain Monte Carlo (MCMC) technique \citep{Verde+2003}.
In our analysis, we use a linear limb-darkening model for the source star.
We apply the linear limb-darkening coefficients as $u_{I}=0.5880$ and $u_{\rm MOA-Red}=0.63445$, which are based on the extinction-corrected source color described in Section \ref{sec:source} and the ATLAS model \citep{Claret+2011}.

\subsection{Heuristic Analysis\label{sec:heuristic}}
We first present a heuristic analysis to predict the binary-lens parameters that produce the anomaly \citep{Gould+1992,Hwang+2018,Skowron+2018}.
This analysis is useful for discussing the origin of the model degeneracy for this event.
The 1L1S modeling yields the parameters $(t_0,t_{\rm eff},t_{\rm E})\sim(6788.0,6.0,35.9)$ days, where $t_0$ is the time when the source approaching closest to the lens, $t_{\rm E}$ is the Einstein radius crossing time, $t_{\rm eff}(\equiv u_0t_{\rm E})$ is the effective event timescale, and $u_0$ is the impact parameter in unit of the angular Einstein radius $\theta_{\rm E}$.
From Figure \ref{fig:lc}, the perturbation is centered at $t_{\rm anom}\sim6770.0$ and then $\tau_{\rm anom}\equiv (t_0-t_{\rm anom})/t_{\rm E}\sim0.45$.
Therefore, if the perturbation is induced by a planet, the source position at the anomaly $u_{\rm anom}$ and the angle between the source trajectory and the binary-axis $\alpha$ \citep{Rattenbury+2002} would be 
\begin{eqnarray}
u_{\rm anom} &=& \sqrt{u_0^2+\tau^2_{\rm anom}} \sim0.53 \nonumber\\
\alpha &=& \tan^{-1}\left(\frac{u_{0}}{\tau_{\rm anom}}\right)\sim18.3^{\circ}. \nonumber
\end{eqnarray}
The projected separation between the lens host and planet in unit of $\theta_{\rm E}$, $s$, can be estimated by $s-s^{-1} = u_{\rm anom}$ and then we approximately estimate $s\sim0.77$ and $1.30$.
This estimation is based on the assumption that the perturbation was induced by the source directly crossing over the planetary caustics.

The remaining binary-lens parameters are $(q,\rho)$, where $q$ is the planet/host mass ratio and $\rho$ is the source angular radius in unit of $\theta_{\rm E}$.
There are no sharp caustic-crossing features in the light curve so that we expect two cases that (1) the source directly passes over the planetary caustic and then fully or partially envelops the caustic (i.e. $q/\rho^2\leq1$) and (2) the source passes near the caustic. 
Here, we consider the case 1 and then estimate $\rho\simeq t_{\rm bump}/2t_{\rm E}=0.04$, where $t_{\rm bump}=3$ days is the duration of the first bump in the anomaly.
Under assumptions that the source is much larger than the Einstein radius of the planet $\theta_{\rm E,p}(\equiv\sqrt{q}\theta_{\rm E})$ and $s>1$, the maximum excess magnification is  $\Delta A=2q/\rho^2$ \citep{Gould+1997}.
In this case, we estimate $q\sim3\times10^{-4}$, where $\Delta A=0.3$.
This result implies that the anomaly was caused by a sub-Jovian-mass-ratio planet.

The heuristic analysis presented above is incomplete because it is based on several assumptions and we do not consider the case in which the source does not directly cross over the caustic.
In order to avoid missing any 2L1S solutions for the event, we conduct a detailed grid search analysis in the next section.
However, as discussed in Section \ref{sec:origin}, this heuristic analysis would help us for understanding what is the origin of the model degeneracy that we will find.

\subsection{Grid Search Analysis}
In order to avoid missing any local 2L1S solutions for the event, we conducted a grid search over the $(q,s,\alpha)$ parameter space where these three parameters are known to strongly affect the magnification pattern.
We set $40\times40\times60=96,000$ grid points distributed at equal intervals over the search ranges of $-6\leq\log{q}\leq0$, $-1\leq\log{s}\leq1$, and $0\leq\alpha<2\pi$, respectively, and then we ran MCMC samplers at each grid point allowing all the parameters to vary except for $(q,s,\alpha)$.
The initial parameters of $(t_0,t_{\rm E},u_0)$ are randomly set to be within the uncertainties of the 1L1S model.
The source radius crossing time, $t_*(=\rho t_{\rm E})$, is also an important parameter for the magnification pattern \citep[e.g.][]{Bennett+2008,Bennett+2014} so that we randomly set an initial value of $\rho$ at each grid point over $10^{-5}<\rho<0.08$.
After that, we refined all the possible models by allowing all the parameters to vary and then excluded models with $\Delta\chi^{2}>100$ compared to the best-fit model.

Our detailed grid search analysis found three different local minima for the 2L1S model.
Here we label the three models in the ranked order of the $\chi^{2}$ goodness-of-fit values as `` model A'', ``model B'', and ``model C''.
The model parameters and $\chi^{2}$ values for each model are summarized in Table \ref{tab:table}.
Figure \ref{fig:models} allows a comparison between the three degenerate models.
As shown in Figure \ref{fig:models}, all the three models fit the light curve perturbation by the sources crossing near or over the planetary caustics, and model A is preferred over model B and C by $\Delta\chi^2=7.5$ and $18.4$, respectively.
We also tried to fit these models including high-order microlensing effects, parallax \citep{Gould2004,Muraki+2011}, xallarap \citep{Poindexter+2005,Miyazaki+2021}, and lens-orbital motion \citep{Skowron+2011}.
However, we found that these effects are not significant for this event and we could not obtain any meaningful constraints from the effects.
{This is probably because the effects are expected to be weak due to the relatively short timescale $t_{\rm E}\sim30$ days and large impact parameter $u_{0}\sim0.2$\footnote{ Moreover, we found the baseline variability of the light curve remains even if we tried to remove it, which can systematically affect our secure measurements on the high-order effects.
Therefore, we focus on the static models.}.}

\begin{deluxetable}{lcccccc}
\tablecaption{Binary-lens Model Parameters\label{tab:table}}
\centering
\tablehead{
Parameters (Units)      &  Model A            & Model B            & Model C 
}
\startdata
$t_0$ (HJD$^\prime-6780$)     & $7.963^{+0.025}_{-0.025}$    & $8.035^{+0.024}_{-0.023}$  & $8.045^{+0.024}_{-0.024}$ \\
$t_{\rm E}$ (day)             & $34.46^{+0.46}_{-0.47}$         & $34.74^{+0.41}_{-0.43}$       & $34.79^{+0.48}_{-0.45}$    \\
$u_0$                            & $0.174^{+0.003}_{-0.003}$       & $0.175^{+0.003}_{-0.003}$     & $0.171^{+0.003}_{-0.003}$  \\
$q\;(10^{-4})$ & $10.34^{+1.31}_{-1.11}$         & $6.56^{+0.61}_{-0.53}$        & $4.51^{+0.87}_{-0.77}$  \\
$s$             & $1.213^{+0.005}_{-0.006}$       & $0.762^{+0.003}_{-0.003}$     & $1.287^{+0.005}_{-0.005}$  \\
$\alpha$ (radian)        & $2.834^{+0.003}_{-0.003}$       & $5.845^{+0.006}_{-0.006}$     & $2.859^{+0.005}_{-0.005}$  \\
$\rho\;(10^{-2})$ & $<2.7\tablenotemark{a}$       & $3.95^{+0.31}_{-0.26}$          & $2.82^{+0.39}_{-0.34}$ \\
$t_*\;({\rm day})$ & $<0.78\tablenotemark{a}$ & $1.37^{+0.10}_{-0.09}$ & $0.98^{+0.13}_{-0.12}$ \\\hline\hline
best-fit $\chi^2$         & 3867.93            & 3875.41            & 3886.31   \\
$\Delta\chi^2$  & $-$                & 7.48              & 18.38     \\\hline   
\enddata
\tablecomments{Medians of the MCMC posterior distributions with uncertainties corresponding to the 68\% credible intervals around the medians.}
\tablenotetext{a}{ $3\sigma$ upper limits.}
\end{deluxetable}

\subsection{Origin of the Degeneracy \label{sec:origin}}
\begin{figure*}
    \centering
    \includegraphics[scale=0.75,angle=270]{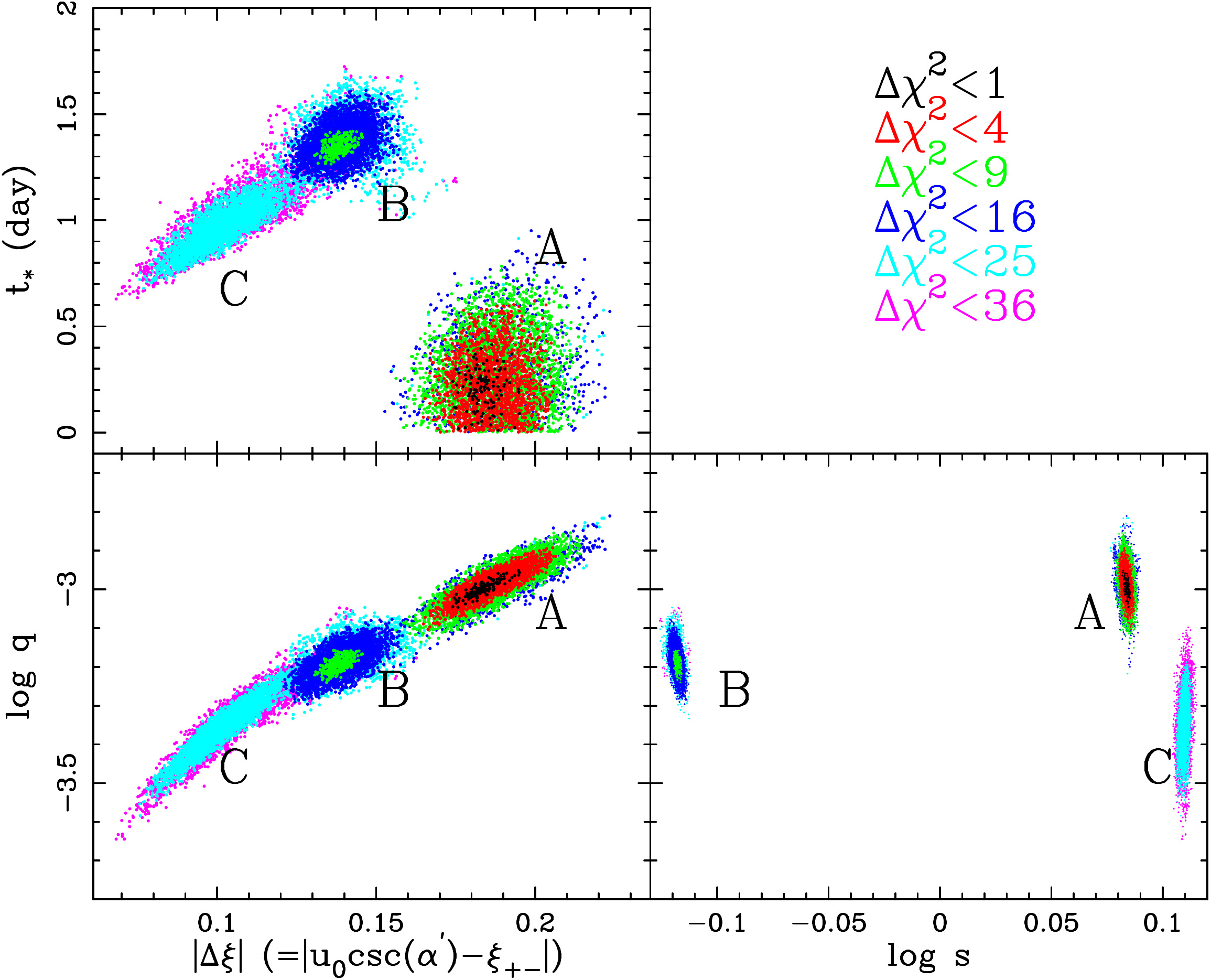}
    \caption{
    The $\Delta\chi^2$ distribution in the MCMC chains on $(|\Delta\xi|,t_*)$, $(|\Delta\xi|,\log q)$, and $(\log s,\log q)$ planes for each model, where the points of black, red, green, blue, cyan, and magenta represent the MCMC links with $\Delta\chi^2<(1,4,9,16,25,36)$, respectively.
    }
    \label{fig:mcmc}
\end{figure*}

Here we discuss the origin of the model degeneracy found in the previous section.
In Section \ref{sec:heuristic}, we estimated $s\sim0.77$ and $1.30$ based on the assumption that the source directly crosses over the planetary caustic.
Models B and C would enter into the group of models we predicted because of the fact that the source crossing over the caustic and the similarity of $s$.
As expected, given there were no sharp caustic-crossing features in the anomaly, the source sizes for the both models are relatively large compared to the caustics, which is shown in the right panels of Figure \ref{fig:models}, and the source sizes are well constrained.
On the other hand, we found model A, with a significantly smaller source size, provides a somewhat better fit.
This model can be considered as the case we did not consider in Section \ref{sec:heuristic}. 
In this case, the source crosses near the planetary caustic to reproduce the smooth perturbation.
To illuminate this discussion, we introduce the parameter \citep{Hwang+2018}, 
\begin{eqnarray}
|\Delta\xi| = |u_0{\rm csc}\alpha^\prime - \xi_{\pm}|,
\end{eqnarray}
where $\alpha^\prime=\alpha\pm n\pi$ is defined to satisfy $\alpha^\prime<\frac{\pi}{2}$ and $\xi_{\pm}=|s-s^{-1}|$.
Hence $|\Delta\xi|$ approximately represents the distance between the center of the source and the caustic when the source crosses the binary-lens axis.
Figure \ref{fig:mcmc} represents the MCMC distribution colored by $\Delta\chi^2$ for each model on $(|\Delta\xi|,t_*)$, $(|\Delta\xi|,\log q)$, and $(\log s,\log q)$ planes.
One sees that the three models seem continuous in the $(|\Delta\xi|,\log q)$ plane.
However, model A is clearly separated from the model B and C on the $(|\Delta\xi|,t_*)$ plane and we note that the models B and C are well-separated on $(\log s,\log q)$ plane.
Therefore, the three models are discrete on the microlensing parameter spaces and thus distinguishable.
Although the model degeneracy for the event is within our expectations from the heuristic analysis and the visual inspection of the light curve, the degeneracy is worth discussing. 

A first degeneracy is between models with $s>1$ and $s<1$ at planetary caustic perturbations, i.e. the degeneracy between models B and C for this event.
In general, the degeneracy could be resolvable because the caustic structures between them are qualitatively quite different and the light-curve features and the timing of the anomalies are different enough to distinguish them.
The reason why the degeneracy occurred in this event would be due to the large source radii that ``wash out'' the magnification patterns and then the both models reproduce similar magnification patterns.
\citet{Gaudi+1997} also predict such a degeneracy, see their Figure 1. 
Recently, \citet[][]{Zang+2021} also encountered the similar degeneracy in a planetary event with $q\sim10^{-5}$ and $\rho\sim10^{-3}$.
A second degeneracy is between models with sources crossing near the caustic and models with large sources crossing over the caustic, i.e. the degeneracy between models A and C for this event.
This degeneracy was also found in \citet{Zang+2020}.
In this degeneracy, one can constrain the source size which is relatively large compared with the caustic and the other can not, which is also demonstrated in \citet{Zang+2020}.
This is qualitatively possible for planetary-caustic events with no sharp caustic-crossing features. 

\section{Physical Properties\label{sec:physical}}
The angular Einstein radius $\theta_{\rm E}$ provides the important mass-distance relation for the lens system: 
\begin{eqnarray}
    M_L = \frac{c^2}{4G}\theta^2_{\rm E}\frac{D_SD_L}{D_S-D_L},
\end{eqnarray}
where $D_S$ and $D_L$ are the distances of the source and the lens from the observer, respectively.
In order to determine $\theta_{\rm E}$, we measure the angular source radius $\theta_{*}\;(\equiv\rho\theta_{\rm E})$ which can be estimated from its color and magnitude \citep{Boyajian+2014}.

\subsection{Source Property \label{sec:source}}
\begin{figure}
    \centering
    \includegraphics[scale=0.35]{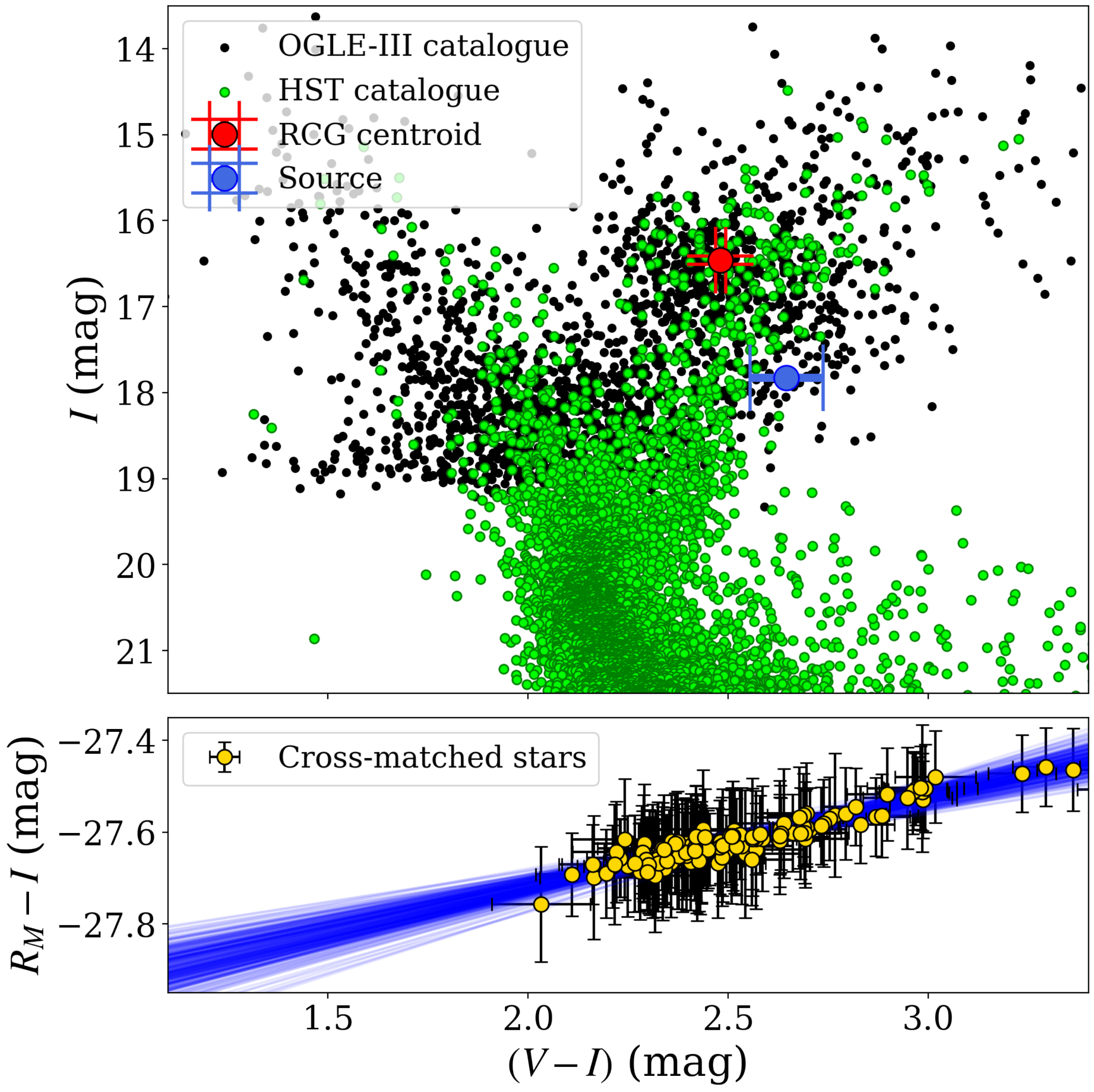}
    \caption{
    (Top panel): The $(V-I,I)$ color magnitude diagram (CMD) in the standard Kron-Cousins $I$ and Johnson $V$ photometric system.
    The positions of the Red Clump (RC) centroid and the source are presented as the red and { blue} circles.
    The black dots indicate stars of the OGLE-III catalogue within $2^\prime$ of the event.
    {The green circles indicate the {\it Hubble Space Telescope (HST)} catalogue \citep{Holzman+1998} whose position is matched using the RC centroid.}
    (Lower panel): The empirical color-color relation between the standard system $(V-I)$ and instrumental color $(R_{\rm MOA}-I)$.
    The blue translucent lines is 300 samples randomly extracted from the MCMC distribution.
    }
    \label{fig:cmd}
\end{figure}

As mentioned in Section \ref{sec:obs}, we do not have any measurements in the $V$-band for this event.
The source magnitudes in the $I$-band ($I_S$) and the MOA-Red band ($R_{M,S}$) are precisely measured from the light curve analysis and these enable us to estimate the source color of $(V-I)$ using an empirical color-color relation between $(R_M-I)$ and $(V-I)$ \citep{Gould+2010,Bennett+2012}. 
We note that the MOA light curve is scaled to match the instrumental MOA-II DOPHOT catalog \citep{Bond+2017} and the OGLE light curve is calibrated to the standard Cousins $I$-band.

We extracted isolated and relatively-bright stars in the OGLE-III catalog which are located within a $2^{\prime}$ circle centered on the event and then cross-referenced the stars in the MOA-II catalog.
We removed stars with colors $(V-I)<2.0$ and then derived the following relation using an MCMC algorithm with 144 cross-matched stars:
\begin{eqnarray}
    R_M-I = (0.196^{+0.026}_{-0.027})(V-I) -28.112^{+0.067}_{-0.068},\nonumber 
\end{eqnarray}
where the parameter uncertainties for the slope and intercept correspond to the 1 sigma credible intervals around the median values in the MCMC stationary distribution.
In the bottom panel of Figure \ref{fig:cmd}, we show 300 relations (blue translucent lines) randomly extracted from the MCMC distribution.
There is a strong correlation between the slope and intercept parameters in the MCMC distribution.
Using the color-color relation, we estimated the source color and magnitude to be $(V-I,I)_S=(2.643,17.829)\pm(0.046, 0.004)$ for model A.
We note that the uncertainty of the source color $(V-I)_S$ largely comes from the uncertainty of the color-color relation.

Figure \ref{fig:cmd} also shows the $(V-I,I)$ color-magnitude diagram (CMD) of the OGLE-III catalog within $2^{\prime}$ of the event which is calibrated to the standard photometric system \citep{Szymanski+2011}.
{
We also plot the CMD of the {\it Hubble Space Telescope (HST)} catalogue \citep{Holzman+1998} and found that the source color is somewhat redder than a sub-giant group in the Galactic bulge.
We estimated the average color of the sub-giant group in the {\it HST} CMD following \citet{Bennett+2008} as $(V-I)_{\rm HST}=2.43\pm0.11$ and confirmed that this is consistent with $(V-I)_S$ at $2\sigma$.
To be consistent with each color, we doubled the nominal error bars on the source color and thus adopted $(V-I)_S=2.643\pm0.093$ in later analysis.
}
Then, we estimated the position of the RC centroid in the CMD to be $(V-I,I)_{\rm RC}=(2.481,16.464)\pm(0.013, 0.050)$.
Comparing $(V-I,I)_{\rm RC}$ with the intrinsic RC color and magnitude $(V-I,I)_{\rm RC,0}=(1.060,14.605)\pm(0.060,0.040)$ \citep{Bensby+2013,Nataf+2013}, we estimated the RC reddening and extinction to be $E(V-I)_{\rm RC}=1.421\pm0.061$ and $A_{I,{\rm RC}}=1.856\pm0.064$, respectively, and then derived the extinction-corrected source color and magnitude $(V-I,I)_{S,0}=(1.223,15.973)\pm(0.111, 0.064)$ on the assumption that the source suffers the same extinction as the RC in the bulge \citep{Yoo+2004}.

Using the intrinsic source color and magnitude and the following empirical relation \citep{Boyajian+2014}:
\begin{eqnarray}
    \log(2\theta_{*}/{\rm mas}) = 0.5014 + 0.4197(V-I)_{S,0} - 0.2I_{S,0},\nonumber 
\end{eqnarray}
we derived the angular source radius $\theta_{*}=3.303\pm0.374\;\mu{\rm as}$ for model A.
We summarize the source properties for each model in Table \ref{tab:source}.

\begin{deluxetable}{lccccccc}
\tablecaption{Source Properties\label{tab:source}}
\centering
\tablehead{
 & Model A & Model B & ModelC
}
\startdata
$(V-I)_{S}$\tablenotemark{a} & $2.643\pm0.093$& $2.675\pm0.096$& $2.648\pm0.093$\\
$I_{S}$ & $17.829\pm0.004$& $17.846\pm0.004$& $17.868\pm0.004$\\
$(V-I)_{S,0}$ & $1.223\pm0.111$& $1.254\pm0.114$& $1.228\pm0.111$\\
$I_{S,0}$ & $15.973\pm0.064$& $15.990\pm0.064$& $16.012\pm0.064$\\
$\theta_{*}\;(\mu{\rm as})$ & $3.303\pm0.374$& $3.378\pm0.390$& $3.261\pm0.370$
\enddata
\tablenotetext{a}{The error bars are extended to be consistent with the average color of a sub-giant group in the {\it HST} CMD.}
\tablecomments{
$(V-I)_{S}$: apparent color; $I_{S}$: apparent magnitude; $(V-I)_{S,0}$: extinction-free color; $I_{S,0}$: extinction-free magnitude; $\theta_{*}$: angular radius.   
}
\end{deluxetable}

\subsection{Angular Einstein Radius}
\begin{figure}
    \centering
    \includegraphics[scale=0.37]{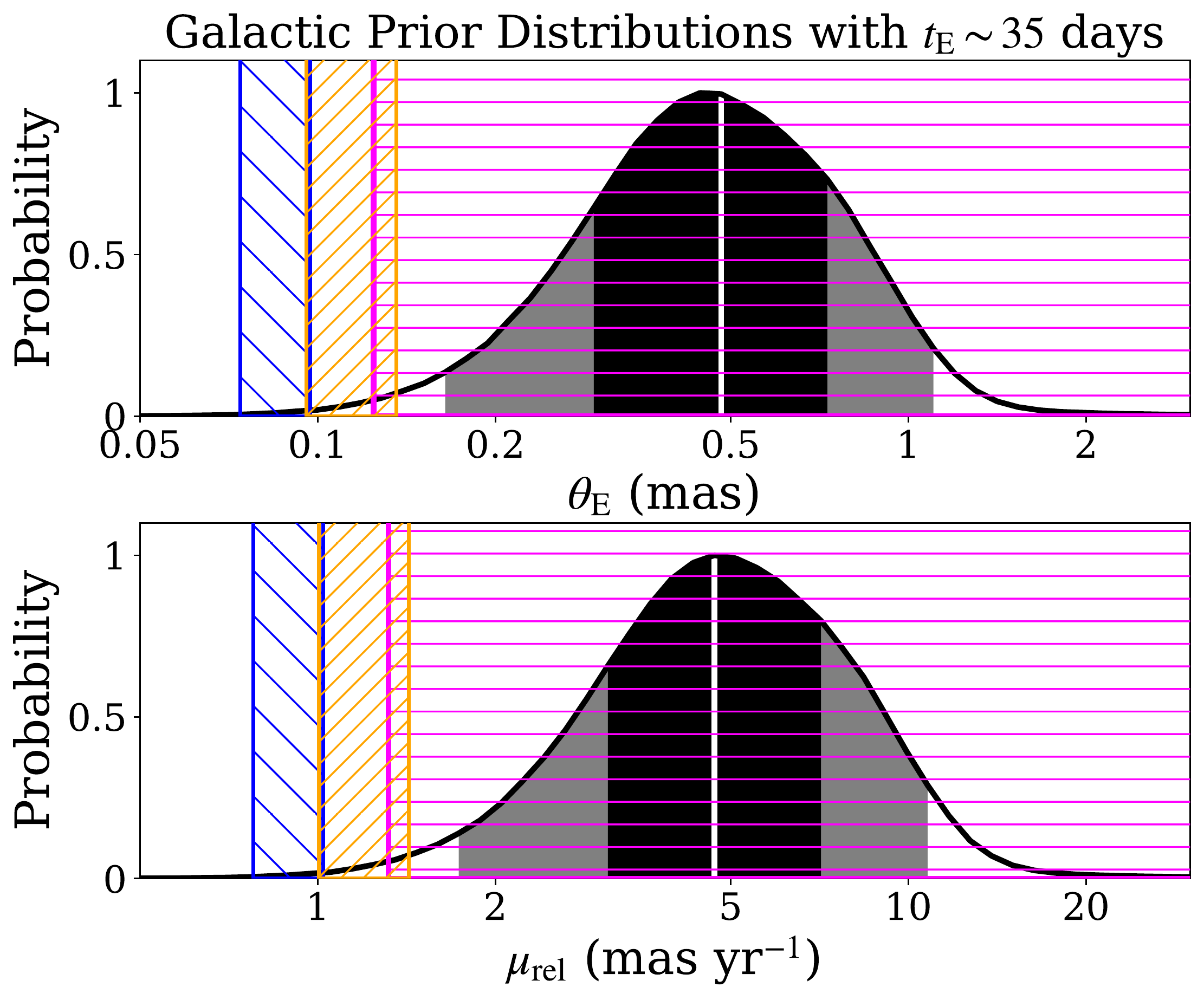}
    \caption{
    Prior probabilities of $\theta_{\rm E}$ and $\mu_{\rm rel}$ derived from a standard Galactic model, where the black and gray regions indicates the $68.3\%\;(1\sigma)$ and $95.4\%\;(2\sigma)$ credible intervals and the vertical white lines indicate the median values.
    Here the prior probabilities are weighted by the observed probability  distribution of $t_{\rm E}\sim35$ day.
    {
    The hatched region colored in magenta represents the 3 $\sigma$ region for model A.
    And also, the regions colored in blue and orange represent the 1 $\sigma$ regions for models B and C, respectively.
    }
    The measured $\theta_{\rm E}$ and $\mu_{\rm rel}$ for models B and C are unreasonably small compared with the prior probabilities.
    }
    \label{fig:Gal_comp}
\end{figure}

Finally, we estimate the angular Einstein radius $\theta_{\rm E}(\equiv\theta_*/\rho)$ and the lens-source relative proper motion $\mu_{\rm rel}(\equiv\theta_{\rm E}/t_{\rm E})$ to be 
\begin{eqnarray}
    \theta_{\rm E}/(\rm mas) = 
    \left\{
        \begin{array}{lc}
            >0.124 & {\rm (for\;Model\;A})\\
            0.086\pm0.012 & {\rm (for\;Model\;B})\\
            0.116\pm0.020 & {\rm (for\;Model\;C}) 
        \end{array}
    \right.\;{\rm,}\;\;\;\nonumber 
\end{eqnarray}
and
\begin{eqnarray}
    \mu_{\rm rel}/(\rm mas/yr) = 
    \left\{
        \begin{array}{lc}
            > 1.316 & {\rm (for\;Model\;A)}\\
            0.90\pm0.12 & {\rm (for\;Model\;B)}\\
            1.21\pm0.21 & {\rm (for\;Model\;C)} 
        \end{array}
    \right., \nonumber
\end{eqnarray}
respectively.
{We note that the limit for model A is the 3 sigma.}

These observed angular Einstein radii for each model provide important constraints on the lens mass $M_L$ and relative lens-source parallax $\pi_{\rm rel}\equiv {\rm au}(D^{-1}_L-D^{-1}_S)$, as 
\begin{eqnarray}
    M_L & = &  \frac{\theta^2_{\rm E}}{\kappa \pi_{\rm rel}} 
    =    \left\{
            \begin{array}{c}
                > 0.063\\
                0.030 \\
                0.055  
            \end{array}
        \right\}
        M_{\odot}\;\left(\frac{0.03\;{\rm mas}}{\pi_{\rm rel}}\right),
\end{eqnarray}
where $\kappa=8.144\;{\rm mas}/M_{\odot}$ and $\pi_{\rm rel}=0.03\;{\rm mas}$ is a typical value for lenses in the Galactic bulge.
This implies that the lenses for models B and C would be sub-stellar objects, which is unlikely for microlensing events with $t_{\rm E}\sim35$ day \citep{Sumi+2011,Mroz+2017}.
To quantitatively demonstrate this, we compared the observed $\theta_{\rm E}$ and $\mu_{\rm rel}$ values with prior probabilities derived from a Bayesian analysis using a Galactic model optimized for use in microlensing studies \citep{Koshimoto+2021} detailed in the Section \ref{sec:Bayesian}.
Figure \ref{fig:Gal_comp} shows the result and indicates that the observed $\theta_{\rm E}$ and $\mu_{\rm rel}$ values for models B and C are $\sim3\sigma$ from the medians.
Therefore, we conclude that models B and C are unlikely  candidates for the best solution.
This is in addition to the fact that they have worse $\chi^2$ values\footnote{Under the assumption of a normal distribution, models B and C can be rejected at significant levels of $\sim0.02$ and $\sim10^{-4}$, respectively.}.
We therefore conclude that model A is the best solution for the event.
However, we should note that the prior probabilities presented in Figure \ref{fig:Gal_comp} do not consider the planet-hosting probability that is likely dependent on properties of the host star, like mass, metallicity, and its Galactic location.
A recent statistical study using 28 planetary events indicates that the dependence on the Galactic location might be small, however \citep{Koshimoto+2021b}.
For completeness, we summarize the final results of models B and C in Appendix \ref{sec:Appendix}.

\subsection{Bayesian Analysis\label{sec:Bayesian}}
\begin{figure*}
    \centering
    \includegraphics[scale=0.52]{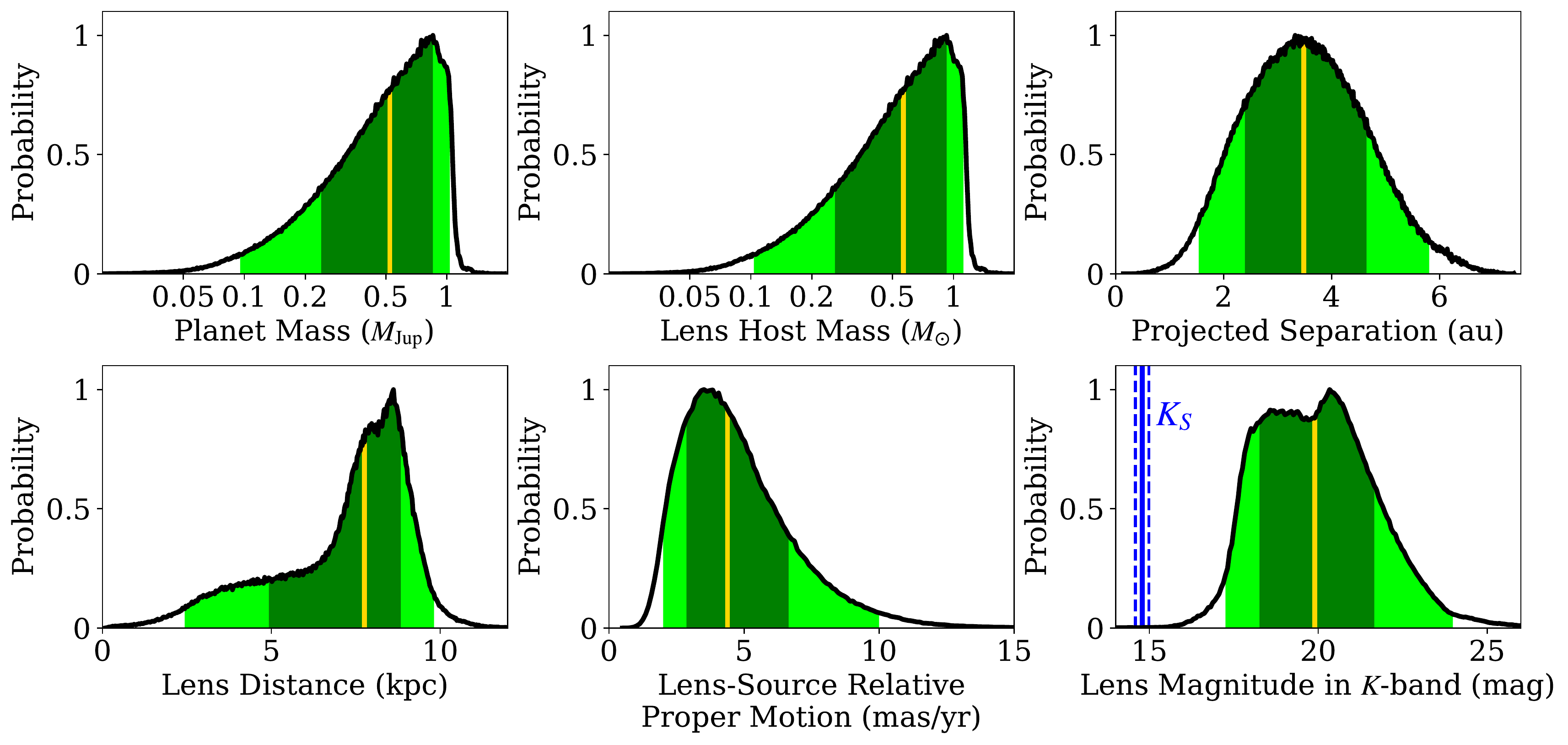}
    \caption{
    Posterior probability distributions of the physical parameters derived from a Bayesian analysis using a Galactic model.
    The green and light green regions indicate the $68.3\%$ ($1\sigma$) and $95.4\%$ ($2\sigma$) credible intervals and the vertical yellow lines indicate the median values.
    We also plot the expected source magnitude $K_S$ in the lower right panel where the dashed lines indicate the $1\sigma$ uncertainties.
    }
    \label{fig:Baysian}
\end{figure*}

\begin{deluxetable}{lcc}
\tablecaption{Physical Parameters\label{tab:lens}}
\centering
\tablehead{
Physical Parameters (Units)      &  Median $\pm1\sigma$
}
\startdata
Planet Mass, $M_P\;(M_{\rm Jup})$& $0.57^{+0.36}_{-0.31}$\\
Lens Host Mass, $M_{\rm Host}\;(M_{\odot})$ & $0.52^{+0.33}_{-0.29}$\\
Projected Semi-major Axis, $a_{\perp}\;({\rm au})$ & $3.49^{+1.17}_{-1.12}$ \\
Distance to the Lens System, $D_L\;({\rm kpc})$ & $7.73^{+1.14}_{-2.84}$ \\
Distance to the Source, $D_S\;({\rm kpc})$ & $10.30^{+3.35}_{-0.68}$\\
Angular Einstein Radius, $\theta_{\rm E}\;({\rm mas})$ & $0.41^{+0.22}_{-0.15}$  \\
Lens-Source Proper Motion, $\mu_{\rm rel}\;({\rm mas/yr})$ & $4.38^{+2.32}_{-1.62}$ \\
Lens Magnitude in $I$-band, $I_L\;({\rm mag})$ & $23.43^{+2.23}_{-2.26}$ \\
Lens Magnitude in $H$-band, $H_L\;({\rm mag})$ & $20.29^{+1.86}_{-1.81}$ \\
Lens Magnitude in $K$-band, $K_L\;({\rm mag})$ & $19.92^{+1.79}_{-1.70}$ \\
Source Magnitude in $H$-band, $H_S\;({\rm mag})$ & $15.04\pm0.20$ \\
Source Magnitude in $K$-band, $K_S\;({\rm mag})$ & $14.79\pm0.20$ 
\enddata
\end{deluxetable}

Because we can not obtain any significant parallax measurements, we can not directly determine the physical properties of the lens system.
Therefore, we conducted a Bayesian analysis in order to quantitatively estimate the probability distribution of the physical properties.
We employed a new parametric Galactic model developed to match recent various observation data toward the Galactic bulge.
We generated $\sim10^6$ artificial microlensing events using the public code {\tt genulens}\footnote{https://github.com/nkoshimoto/genulens} \citep{Koshimoto&Ranc2021}.
We then derived the posterior probability distribution weighting by the observed likelihood distributions of $t_{\rm E}$, $\theta_{\rm E}$, $I_S$, and $(V-I)_S$.
The result of the Bayesian analysis is shown in Figure \ref{fig:Baysian} and the physical parameters derived are summarized in Table \ref{tab:lens}.  
The analysis indicates that the lens system comprises of a sub-Jovian-mass planet orbiting an M-dwarf star near the Galactic bulge.

\subsection{Future Follow-up Observation}
Future follow-up observations with high-angular-resolution might be able to give an additional mass-distance relation derived from the lens flux.
This can resolve the large uncertainties of the lens physical parameters \citep{Bennett+2015,Bhattacharya+2018,Bennett+2020,Bhattacharya+2020,Vandorou+2020,Terry+2021}.
For this, we also estimate the apparent magnitude of the lens brightness in the Bayesian analysis.
Here we model the extinction in front of the lens \citep{Bennett+2015} as 
\begin{eqnarray}
A_L = \frac{1-e^{-D_L/(0.1{\rm kpc})\sin{|b|}}}{1-e^{-D_S/(0.1{\rm kpc})\sin{|b|}}}A_S
\end{eqnarray}
where $A_L$ and $A_S$ are the extinction values for the lens and source systems, respectively.  
We used the mass-luminosity and color-color relations \citep{Henry+1993,Kenyon+1995,Kroupa+1993} and the extinction law \citep{Nishiyama+2009} to derive the apparent magnitudes of the lens and the source in different pass-bands.
The result is summarized in Table \ref{tab:lens}.
It is expected that the source is $\sim5.1$ mag brighter than the lens in $K$-band and the angular separation between the source and the lens will be $\sim35$ mas in 2022.
Owing to such a high contrast, several more years might need to pass in order to resolve the lens position and detect the lens flux.
However, observing the lens not only can resolve the uncertainties of the physical parameters, it can also break the model degeneracy because each degenerate model indicates different lens-source relative proper motions $\mu_{\rm rel}$.
Therefore, it would be worth conducting these future follow-up observations.

\section{Summary \& Discussion\label{sec:discussion}}

We present the analysis of the planetary microlensing event OGLE-2014-BLG-0319.
We find three possible models with different mass ratios $q=(10.3,6.6,4.5)\times10^{-4}$ and lens-source relative proper motions $\mu_{\rm rel}=(>2.50,0.90,1.26)$ mas/yr, respectively.
We rule out the last two models with small $\mu_{\rm rel}$ values considering the Galactic prior probability.
Finally, we conduct a Bayesian analysis using a Galactic model and estimate the lens system consists of a sub-Jupiter-mass planet orbiting an M-dwarf near the Galactic bulge.

\subsection{Degeneracy of Mass Ratios and Proper Motions}
In general, the mass-ratio $q$ is approximately determined from the duration of planetary perturbation relative to $t_{\rm E}$. 
If the source size is large relative to the Einstein radius of the planet, then the duration of the perturbation becomes the source crossing time.
\citet{Gaudi+1997} predict a continuous degeneracy in such a case: $\mu_{\rm rel}\rho q^{-1/2}=const$.
At a given $\mu_{\rm rel}$, the $q^{-1/2}$ and $\rho$ can be degenerate.
They also propose that the low-mass solutions with low $\mu_{\rm rel}$ values could be \textit{a priori} ruled out.
For the event OGLE-2014-BLG-0319, we can rule out models B and C by comparing the measured values and the Galactic prior probabilities for $\theta_{\rm E}$ and $\mu_{\rm rel}$.
This approach can be made because the observed values are unlikely at a given $t_{\rm E}\sim35$ days\footnote{
Interestingly, \citet{Han+2020} reported a discovery of a planetary microlensing event with $t_{\rm E}=45$ days and $\mu_{\rm rel}\sim0.79$ mas/yr.
Note that they confirmed that there are no competing models with the final result.
}.
However, we expect this would be more difficult for short-timescale (low-mass lens) events because $\rho\propto t^{-1}_{\rm E}$ and $t_{\rm E}\propto\sqrt{M_{L}}$.
{For example, a Bayesian analysis with $t_{\rm E}\sim5$ days provides high prior probabilities at $\theta_{\rm E}\sim0.1$ mas and $\mu_{\rm rel}\sim1$ mas/yr}.

So far, there have been several reported short timescale events which suffered from similar model degeneracies between different $q$ and $\mu_{\rm rel}$ values \citep[e.g.,][]{Miyazaki+2018,Zhang+2020}.
These events have several degenerate solutions with $\rho \geq 0.01$ and $t_{\rm E} \leq 10$ days and they can not be ruled out by Galactic priors.
As discussed in \citet{Zhang+2020}, the discovery rate of short timescale events with planetary (short-lived) perturbations has much increased since current second-generation microlensing surveys started.
Hence, we can now explore planets around low-mass dwarfs, brown dwarfs, and even planetary-mass objects.
However, the degeneracy would be more common for short timescale events and thus might have large impacts on the estimation for the frequency of such planets. 

\subsection{Detection Efficiency Dependent on Source Size}
\begin{figure}
    \centering
    \includegraphics[scale=0.4]{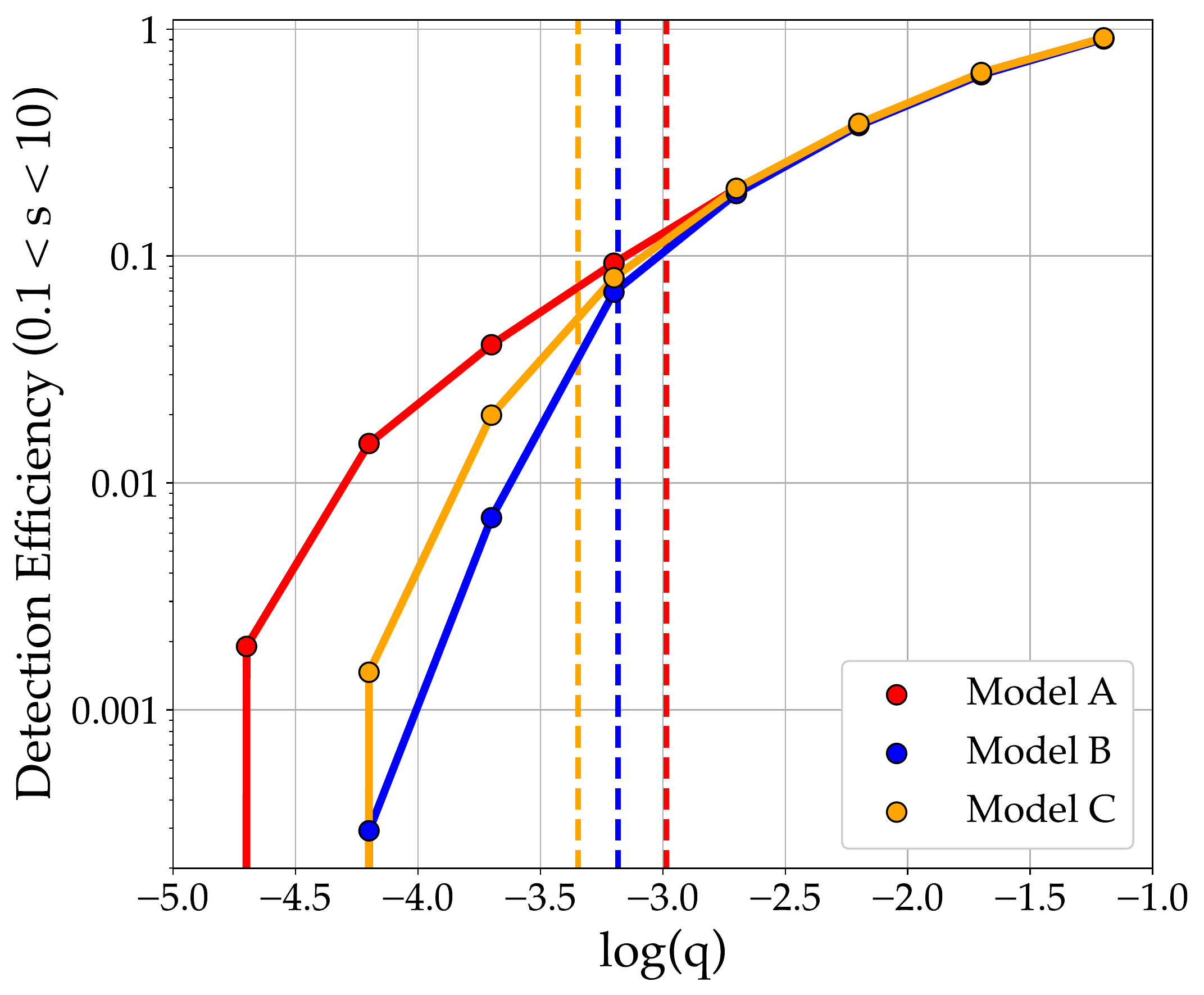}
    \caption{
    Detection efficiencies for each degenerate model as a function of mass-ratio $q$.
    The vertical dashed lines represents the observed $q$ for each model.
    {For model A, we used the best-fit value of $\rho=7.0\times10^{-3}$ for calculation.}
    }
    \label{fig:detection_eff}
\end{figure}

Furthermore, estimating different $\rho$ values can lead to different detection efficiencies of planets for each event.
Figure \ref{fig:detection_eff} shows the planet detection efficiencies for each model as a function of $\log q$, in which the dashed vertical lines represent the observed $q$ for each model ({Details of the calculation for the detection efficiency are similar with \citealp{Suzuki+2016} and described in Section 4 of \citealp{Suzuki+2016}.}).
For this event, the different $\rho$ estimation for each model does not strongly affect the efficiency at $\log q \sim-3$.
However, the effect would be more serious at $\log q \ll-3$  where the detection efficiencies are different by an order of magnitude, as shown in Figure \ref{fig:detection_eff}.
This indicates that the detection efficiency of low mass-ratio planets around low-mass objects would largely be dependent on $\rho\;(\equiv \theta_*/\mu_{\rm rel}t_{\rm E})$ and calls our attention to some assumptions to estimate the detection efficiency.
{Recent planet discoveries with $q<2\times10^{-5}$ have also shown that smaller $\rho$ is more sensitive to low-mass planets; $\rho\sim5\times10^{-4}$ \citep[KMT-2018-BLG-0029,][]{Gould+2020}, $\rho\sim3\times10^{-4}$ \citep[OGLE-2019-BLG-0960,][]{Yee+2021}, $\rho\sim2\times10^{-3}$ \citep[OGLE-2019-BLG-1053,][]{Zang+2021b}, and $\rho\sim6\times10^{-4}$ \citep[KMT-2020-BLG-0414,][]{Zang+2021}.}

For example, \citet{Suzuki+2016} derived $\rho$ values assuming that single-lens events have the similar $\mu_{\rm rel}$ distribution to their planetary sample of $\mu_{\rm rel}=5.74^{+2.94}_{-2.84}\;{\rm mas\;yr^{-1}}$, and then calculated the detection efficiency for each event\footnote{Note that they calculated the detection efficiencies for each single-lens event using a value of $\mu_{\rm rel}=5.6$ mas/yr so that their final result does not include the uncertainties of $\mu_{\rm rel}$.
However, they showed it hardly affects the final result.}.
They demonstrated that the efficiency does not significantly depend on the derived $\rho$ values within the $1\sigma$ uncertainties (See Figure 7 of \citealp{Suzuki+2016}).
Moreover, the $\mu_{\rm rel}$ distribution they assumed is also consistent with the prediction of a Galactic model.
However, it is unclear whether the assumption holds up in the low-mass regime because the Galactic model for low-mass objects like velocity distribution is still much less clear and might be different with those in stellar-mass regime \citep{Sajadian+2021}.
Considering the effect could be important to estimate the frequency of planets around low-mass dwarfs, brown dwarfs \citep{Lingam+2020}, and even moons around free-floating planets \citep{Bennett+2014}.
This would effectively be investigated by the Roman Galactic Exoplanet Survey \citep[RGES;][]{Penny+2019,Johnson+2020} which has much better sensitivity to short-timescale events than current microlensing surveys.

\begin{acknowledgments}
We would appreciate the anonymous referee for helping to improve the paper.
We acknowledge the following support: Work by SM is supported by JSPS KAKENHI Grant Number 21J11296.
The work of DPB., NK, CR, YH, and AB was supported supported by NASA under award number 80GSFC17M0002.
CR is supported by the ANR project COLD-WORLDS of the French {\itshape Agence Nationale de la Recherche} with the reference ANR-18-CE31-0002.
The MOA project is supported by JSPS KAKENHI grant No. 19KK0082 and 20H04754.
\end{acknowledgments}

\appendix
\section{Physical Parameters for Models B and C \label{sec:Appendix}}
For completeness, we present Table \ref{tab:lens2} that represents the physical parameters for models B and C, which is derived by the Bayesian analysis presented in Section \ref{sec:Bayesian}.
\begin{deluxetable*}{lcc}
\tablecaption{Physical Parameters for Models B and C\label{tab:lens2}}
\centering
\tablehead{
Physical Parameters (Units)      &  Model B & Model C
}
\startdata
Planet Mass, $M_P\;(M_{\oplus})$& $17.05^{+25.71}_{-9.45}$ & $17.14^{+24.35}_{-8.48}$\\
Lens Host Mass, $M_{\rm Host}\;(M_{\odot})$ & $0.08^{+0.12}_{-0.04}$ & $0.11^{+0.16}_{-0.06}$\\
Projected Semi-major Axis, $a_{\perp}\;({\rm au})$ & $0.64^{+0.10}_{-0.09}$ & $1.45^{+0.27}_{-0.24}$ \\
Distance to the Lens System, $D_L\;({\rm kpc})$ & $8.96^{+0.55}_{-0.63}$ & $8.92^{+0.57}_{-0.71}$\\
Distance to the Source, $D_S\;({\rm kpc})$ & $10.12^{+1.85}_{-0.63}$ & $10.29^{+1.67}_{-0.65}$\\
Angular Einstein Radius, $\theta_{\rm E}\;({\rm mas})$ & $0.10^{+0.01}_{-0.02}$ & $0.13^{+0.02}_{-0.02}$  \\
Lens-Source Proper Motion, $\mu_{\rm rel}\;({\rm mas/yr})$ &  $0.97^{+0.12}_{-0.12}$ & $1.32^{+0.20}_{-0.19}$\\
Lens Magnitude in $I$-band, $I_L\;({\rm mag})$ & $28.29^{+2.68}_{-2.12}$ & $27.83^{+2.30}_{-2.04}$\\
Lens Magnitude in $H$-band, $H_L\;({\rm mag})$ & $26.61^{+6.16}_{-3.27}$ & $24.51^{+5.89}_{-1.92}$\\
Lens Magnitude in $K$-band, $K_L\;({\rm mag})$ & $26.07^{+7.66}_{-3.21}$ & $24.02^{+6.31}_{-1.89}$\\
Source Magnitude in $H$-band, $H_S\;({\rm mag})$ & $15.04\pm0.20$ & $15.08\pm0.20$ \\
Source Magnitude in $K$-band, $K_S\;({\rm mag})$ & $14.78\pm0.20$ & $14.82\pm0.20$
\enddata
\end{deluxetable*}

\bibliographystyle{aasjournal}
\bibliography{reference.bib}

\begin{thebibliography}{}
\expandafter\ifx\csname natexlab\endcsname\relax\def\natexlab#1{#1}\fi
\providecommand{\url}[1]{\href{#1}{#1}}
\providecommand{\dodoi}[1]{doi:~\href{http://doi.org/#1}{\nolinkurl{#1}}}
\providecommand{\doeprint}[1]{\href{http://ascl.net/#1}{\nolinkurl{http://ascl.net/#1}}}
\providecommand{\doarXiv}[1]{\href{https://arxiv.org/abs/#1}{\nolinkurl{https://arxiv.org/abs/#1}}}

\bibitem[{{Akeson} {et~al.}(2013){Akeson}, {Chen}, {Ciardi}, {Crane}, {Good},
  {Harbut}, {Jackson}, {Kane}, {Laity}, {Leifer}, {Lynn}, {McElroy}, {Papin},
  {Plavchan}, {Ram{\'\i}rez}, {Rey}, {von Braun}, {Wittman}, {Abajian}, {Ali},
  {Beichman}, {Beekley}, {Berriman}, {Berukoff}, {Bryden}, {Chan}, {Groom},
  {Lau}, {Payne}, {Regelson}, {Saucedo}, {Schmitz}, {Stauffer}, {Wyatt}, \&
  {Zhang}}]{Akeson+2013}
{Akeson}, R.~L., {Chen}, X., {Ciardi}, D., {et~al.} 2013, \pasp, 125, 989,
  \dodoi{10.1086/672273}

\bibitem[{{Alard} \& {Lupton}(1998)}]{Alard+1998}
{Alard}, C., \& {Lupton}, R.~H. 1998, \apj, 503, 325, \dodoi{10.1086/305984}

\bibitem[{{Bennett}(2010)}]{Bennett2010}
{Bennett}, D.~P. 2010, \apj, 716, 1408, \dodoi{10.1088/0004-637X/716/2/1408}

\bibitem[{{Bennett} {et~al.}(2021){Bennett}, {Ranc}, \&
  {Fernandes}}]{Bennett+2021}
{Bennett}, D.~P., {Ranc}, C., \& {Fernandes}, R.~B. 2021, arXiv e-prints,
  arXiv:2104.05713.
\newblock \doarXiv{2104.05713}

\bibitem[{{Bennett} \& {Rhie}(1996)}]{Bennett+1996}
{Bennett}, D.~P., \& {Rhie}, S.~H. 1996, \apj, 472, 660, \dodoi{10.1086/178096}

\bibitem[{{Bennett} {et~al.}(2008){Bennett}, {Bond}, {Udalski}, {Sumi}, {Abe},
  {Fukui}, {Furusawa}, {Hearnshaw}, {Holderness}, {Itow}, {Kamiya}, {Korpela},
  {Kilmartin}, {Lin}, {Ling}, {Masuda}, {Matsubara}, {Miyake}, {Muraki},
  {Nagaya}, {Okumura}, {Ohnishi}, {Perrott}, {Rattenbury}, {Sako}, {Saito},
  {Sato}, {Skuljan}, {Sullivan}, {Sweatman}, {Tristram}, {Yock}, {Kubiak},
  {Szyma{\'n}ski}, {Pietrzy{\'n}ski}, {Soszy{\'n}ski}, {Szewczyk},
  {Wyrzykowski}, {Ulaczyk}, {Batista}, {Beaulieu}, {Brillant}, {Cassan},
  {Fouqu{\'e}}, {Kervella}, {Kubas}, \& {Marquette}}]{Bennett+2008}
{Bennett}, D.~P., {Bond}, I.~A., {Udalski}, A., {et~al.} 2008, \apj, 684, 663,
  \dodoi{10.1086/589940}

\bibitem[{{Bennett} {et~al.}(2012){Bennett}, {Sumi}, {Bond}, {Kamiya}, {Abe},
  {Botzler}, {Fukui}, {Furusawa}, {Itow}, {Korpela}, {Kilmartin}, {Ling},
  {Masuda}, {Matsubara}, {Miyake}, {Muraki}, {Ohnishi}, {Rattenbury}, {Saito},
  {Sullivan}, {Suzuki}, {Sweatman}, {Tristram}, {Wada}, {Yock}, \& {MOA
  Collaboration}}]{Bennett+2012}
{Bennett}, D.~P., {Sumi}, T., {Bond}, I.~A., {et~al.} 2012, \apj, 757, 119,
  \dodoi{10.1088/0004-637X/757/2/119}

\bibitem[{{Bennett} {et~al.}(2014){Bennett}, {Batista}, {Bond}, {Bennett},
  {Suzuki}, {Beaulieu}, {Udalski}, {Donatowicz}, {Bozza}, {Abe}, {Botzler},
  {Freeman}, {Fukunaga}, {Fukui}, {Itow}, {Koshimoto}, {Ling}, {Masuda},
  {Matsubara}, {Muraki}, {Namba}, {Ohnishi}, {Rattenbury}, {Saito}, {Sullivan},
  {Sumi}, {Sweatman}, {Tristram}, {Tsurumi}, {Wada}, {Yock}, {MOA
  Collaboration}, {Albrow}, {Bachelet}, {Brillant}, {Caldwell}, {Cassan},
  {Cole}, {Corrales}, {Coutures}, {Dieters}, {Dominis Prester}, {Fouqu{\'e}},
  {Greenhill}, {Horne}, {Koo}, {Kubas}, {Marquette}, {Martin}, {Menzies},
  {Sahu}, {Wambsganss}, {Williams}, {Zub}, {PLANET Collaboration}, {Choi},
  {DePoy}, {Dong}, {Gaudi}, {Gould}, {Han}, {Henderson}, {McGregor}, {Lee},
  {Pogge}, {Shin}, {Yee}, {{\ensuremath{\mu}}FUN Collaboration},
  {Szyma{\'n}ski}, {Skowron}, {Poleski}, {Koz{\l}owski}, {Wyrzykowski},
  {Kubiak}, {Pietrukowicz}, {Pietrzy{\'n}ski}, {Soszy{\'n}ski}, {Ulaczyk},
  {OGLE Collaboration}, {Tsapras}, {Street}, {Dominik}, {Bramich}, {Browne},
  {Hundertmark}, {Kains}, {Snodgrass}, {Steele}, {RoboNet Collaboration},
  {Dekany}, {Gonzalez}, {Heyrovsk{\'y}}, {Kandori}, {Kerins}, {Lucas},
  {Minniti}, {Nagayama}, {Rejkuba}, {Robin}, \& {Saito}}]{Bennett+2014}
{Bennett}, D.~P., {Batista}, V., {Bond}, I.~A., {et~al.} 2014, \apj, 785, 155,
  \dodoi{10.1088/0004-637X/785/2/155}

\bibitem[{{Bennett} {et~al.}(2015){Bennett}, {Bhattacharya}, {Anderson},
  {Bond}, {Anderson}, {Barry}, {Batista}, {Beaulieu}, {DePoy}, {Dong}, {Gaudi},
  {Gilbert}, {Gould}, {Pfeifle}, {Pogge}, {Suzuki}, {Terry}, \&
  {Udalski}}]{Bennett+2015}
{Bennett}, D.~P., {Bhattacharya}, A., {Anderson}, J., {et~al.} 2015, \apj, 808,
  169, \dodoi{10.1088/0004-637X/808/2/169}

\bibitem[{{Bennett} {et~al.}(2020){Bennett}, {Bhattacharya}, {Beaulieu},
  {Blackman}, {Vandorou}, {Terry}, {Cole}, {Henderson}, {Koshimoto}, {Lu},
  {Baptiste Marquette}, {Ranc}, \& {Udalski}}]{Bennett+2020}
{Bennett}, D.~P., {Bhattacharya}, A., {Beaulieu}, J.-P., {et~al.} 2020, \aj,
  159, 68, \dodoi{10.3847/1538-3881/ab6212}

\bibitem[{{Bensby} {et~al.}(2013){Bensby}, {Yee}, {Feltzing}, {Johnson},
  {Gould}, {Cohen}, {Asplund}, {Mel{\'e}ndez}, {Lucatello}, {Han}, {Thompson},
  {Gal-Yam}, {Udalski}, {Bennett}, {Bond}, {Kohei}, {Sumi}, {Suzuki}, {Suzuki},
  {Takino}, {Tristram}, {Yamai}, \& {Yonehara}}]{Bensby+2013}
{Bensby}, T., {Yee}, J.~C., {Feltzing}, S., {et~al.} 2013, \aap, 549, A147,
  \dodoi{10.1051/0004-6361/201220678}

\bibitem[{{Bhattacharya} {et~al.}(2018){Bhattacharya}, {Beaulieu}, {Bennett},
  {Anderson}, {Koshimoto}, {Lu}, {Batista}, {Blackman}, {Bond}, {Fukui},
  {Henderson}, {Hirao}, {Marquette}, {Mroz}, {Ranc}, \&
  {Udalski}}]{Bhattacharya+2018}
{Bhattacharya}, A., {Beaulieu}, J.~P., {Bennett}, D.~P., {et~al.} 2018, \aj,
  156, 289, \dodoi{10.3847/1538-3881/aaed46}

\bibitem[{{Bhattacharya} {et~al.}(2020){Bhattacharya}, {Bennett}, {Beaulieu},
  {Bond}, {Koshimoto}, {Lu}, {Blackman}, {Vandorou}, {Terry}, {Batista},
  {Marquette}, {Cole}, {Fukui}, \& {Henderson}}]{Bhattacharya+2020}
{Bhattacharya}, A., {Bennett}, D.~P., {Beaulieu}, J.~P., {et~al.} 2020, arXiv
  e-prints, arXiv:2009.02329.
\newblock \doarXiv{2009.02329}

\bibitem[{{Bond} {et~al.}(2001){Bond}, {Abe}, {Dodd}, {Hearnshaw}, {Honda},
  {Jugaku}, {Kilmartin}, {Marles}, {Masuda}, {Matsubara}, {Muraki}, {Nakamura},
  {Nankivell}, {Noda}, {Noguchi}, {Ohnishi}, {Rattenbury}, {Reid}, {Saito},
  {Sato}, {Sekiguchi}, {Skuljan}, {Sullivan}, {Sumi}, {Takeuti}, {Watase},
  {Wilkinson}, {Yamada}, {Yanagisawa}, \& {Yock}}]{Bond+2001}
{Bond}, I.~A., {Abe}, F., {Dodd}, R.~J., {et~al.} 2001, \mnras, 327, 868,
  \dodoi{10.1046/j.1365-8711.2001.04776.x}

\bibitem[{{Bond} {et~al.}(2017){Bond}, {Bennett}, {Sumi}, {Udalski}, {Suzuki},
  {Rattenbury}, {Bozza}, {Koshimoto}, {Abe}, {Asakura}, {Barry},
  {Bhattacharya}, {Donachie}, {Evans}, {Fukui}, {Hirao}, {Itow}, {Li}, {Ling},
  {Masuda}, {Matsubara}, {Muraki}, {Nagakane}, {Ohnishi}, {Ranc}, {Saito},
  {Sharan}, {Sullivan}, {Tristram}, {Yamada}, {Yamada}, {Yonehara}, {Skowron},
  {Szyma{\'n}ski}, {Poleski}, {Mr{\'o}z}, {Soszy{\'n}ski}, {Pietrukowicz},
  {Koz{\l}owski}, {Ulaczyk}, \& {Pawlak}}]{Bond+2017}
{Bond}, I.~A., {Bennett}, D.~P., {Sumi}, T., {et~al.} 2017, \mnras, 469, 2434,
  \dodoi{10.1093/mnras/stx1049}

\bibitem[{{Boyajian} {et~al.}(2014){Boyajian}, {van Belle}, \& {von
  Braun}}]{Boyajian+2014}
{Boyajian}, T.~S., {van Belle}, G., \& {von Braun}, K. 2014, \aj, 147, 47,
  \dodoi{10.1088/0004-6256/147/3/47}

\bibitem[{{Butler} {et~al.}(2006){Butler}, {Wright}, {Marcy}, {Fischer},
  {Vogt}, {Tinney}, {Jones}, {Carter}, {Johnson}, {McCarthy}, \&
  {Penny}}]{Butler+2006}
{Butler}, R.~P., {Wright}, J.~T., {Marcy}, G.~W., {et~al.} 2006, \apj, 646,
  505, \dodoi{10.1086/504701}

\bibitem[{{Charbonneau} {et~al.}(2000){Charbonneau}, {Brown}, {Latham}, \&
  {Mayor}}]{Charbonneau+2000}
{Charbonneau}, D., {Brown}, T.~M., {Latham}, D.~W., \& {Mayor}, M. 2000, \apjl,
  529, L45, \dodoi{10.1086/312457}

\bibitem[{{Claret} \& {Bloemen}(2011)}]{Claret+2011}
{Claret}, A., \& {Bloemen}, S. 2011, \aap, 529, A75,
  \dodoi{10.1051/0004-6361/201116451}

\bibitem[{{Cumming} {et~al.}(2008){Cumming}, {Butler}, {Marcy}, {Vogt},
  {Wright}, \& {Fischer}}]{Cumming+2008}
{Cumming}, A., {Butler}, R.~P., {Marcy}, G.~W., {et~al.} 2008, \pasp, 120, 531,
  \dodoi{10.1086/588487}

\bibitem[{{Fressin} {et~al.}(2013){Fressin}, {Torres}, {Charbonneau}, {Bryson},
  {Christiansen}, {Dressing}, {Jenkins}, {Walkowicz}, \&
  {Batalha}}]{Fressin+2013}
{Fressin}, F., {Torres}, G., {Charbonneau}, D., {et~al.} 2013, \apj, 766, 81,
  \dodoi{10.1088/0004-637X/766/2/81}

\bibitem[{{Gaudi}(1998)}]{Gaudi1998}
{Gaudi}, B.~S. 1998, \apj, 506, 533, \dodoi{10.1086/306256}

\bibitem[{{Gaudi} \& {Gould}(1997)}]{Gaudi+1997}
{Gaudi}, B.~S., \& {Gould}, A. 1997, \apj, 486, 85, \dodoi{10.1086/304491}

\bibitem[{{Gould}(2004)}]{Gould2004}
{Gould}, A. 2004, \apj, 606, 319, \dodoi{10.1086/382782}

\bibitem[{{Gould} {et~al.}(2010){Gould}, {Dong}, {Bennett}, {Bond}, {Udalski},
  \& {Kozlowski}}]{Gould+2010}
{Gould}, A., {Dong}, S., {Bennett}, D.~P., {et~al.} 2010, \apj, 710, 1800,
  \dodoi{10.1088/0004-637X/710/2/1800}

\bibitem[{{Gould} \& {Gaucherel}(1997)}]{Gould+1997}
{Gould}, A., \& {Gaucherel}, C. 1997, \apj, 477, 580, \dodoi{10.1086/303751}

\bibitem[{{Gould} \& {Loeb}(1992)}]{Gould+1992}
{Gould}, A., \& {Loeb}, A. 1992, \apj, 396, 104, \dodoi{10.1086/171700}

\bibitem[{{Gould} {et~al.}(2020){Gould}, {Ryu}, {Calchi Novati}, {Zang},
  {Albrow}, {Chung}, {Han}, {Hwang}, {Jung}, {Shin}, {Shvartzvald}, {Yee},
  {Cha}, {Kim}, {Kim}, {Kim}, {Lee}, {Lee}, {Lee}, {Park}, {Pogge}, {Beichman},
  {Bryden}, {Carey}, {Gaudi}, {Henderson}, {Zhu}, {Fouque}, {Penny}, {Petric},
  {Burdullis}, \& {Mao}}]{Gould+2020}
{Gould}, A., {Ryu}, Y.-H., {Calchi Novati}, S., {et~al.} 2020, Journal of
  Korean Astronomical Society, 53, 9, \dodoi{10.5303/JKAS.2020.53.1.9}

\bibitem[{{Han} {et~al.}(2018){Han}, {Bond}, {Gould}, {Albrow}, {Chung},
  {Jung}, {Hwang}, {Lee}, {Ryu}, {Shin}, {Shvartzvald}, {Yee}, {Cha}, {Kim},
  {Kim}, {Kim}, {Lee}, {Lee}, {Park}, {Pogge}, {Kim}, {KMTNet Collaboration},
  {Abe}, {Barry}, {Bennett}, {Bhattacharya}, {Donachie}, {Fukui}, {Hirao},
  {Itow}, {Kawasaki}, {Kondo}, {Koshimoto}, {Li}, {Matsubara}, {Muraki},
  {Miyazaki}, {Nagakane}, {Ranc}, {Rattenbury}, {Suematsu}, {Sullivan}, {Sumi},
  {Suzuki}, {Tristram}, {Yonehara}, \& {MOA Collaboration}}]{Han+2018}
{Han}, C., {Bond}, I.~A., {Gould}, A., {et~al.} 2018, \aj, 156, 226,
  \dodoi{10.3847/1538-3881/aae38e}

\bibitem[{{Han} {et~al.}(2020){Han}, {Udalsk}, {Gould}, {Albrow}, {Chung},
  {Hwang}, {Jung}, {Lee}, {Ryu}, {Shin}, {Shvartzvald}, {Yee}, {Zang}, {Cha},
  {Kim}, {Kim}, {Kim}, {Lee}, {Lee}, {Park}, {Pogge}, {Jee}, {Kim}, {Kim},
  {Kim}, {Mr{\'o}z}, {Szyma{\'n}ski}, {Skowron}, {Poleski}, {Soszy{\'n}ski},
  {Pietrukowicz}, {Koz{\l}owski}, \& {Ulaczyk}}]{Han+2020}
{Han}, C., {Udalsk}, A., {Gould}, A., {et~al.} 2020, \aj, 159, 91,
  \dodoi{10.3847/1538-3881/ab6a9f}

\bibitem[{{Henry} \& {McCarthy}(1993)}]{Henry+1993}
{Henry}, T.~J., \& {McCarthy}, Donald~W., J. 1993, \aj, 106, 773,
  \dodoi{10.1086/116685}

\bibitem[{{Holtzman} {et~al.}(1998){Holtzman}, {Watson}, {Baum}, {Grillmair},
  {Groth}, {Light}, {Lynds}, \& {O'Neil}}]{Holzman+1998}
{Holtzman}, J.~A., {Watson}, A.~M., {Baum}, W.~A., {et~al.} 1998, \aj, 115,
  1946, \dodoi{10.1086/300336}

\bibitem[{{Howard} {et~al.}(2012){Howard}, {Marcy}, {Bryson}, {Jenkins},
  {Rowe}, {Batalha}, {Borucki}, {Koch}, {Dunham}, {Gautier}, {Van Cleve},
  {Cochran}, {Latham}, {Lissauer}, {Torres}, {Brown}, {Gilliland}, {Buchhave},
  {Caldwell}, {Christensen-Dalsgaard}, {Ciardi}, {Fressin}, {Haas}, {Howell},
  {Kjeldsen}, {Seager}, {Rogers}, {Sasselov}, {Steffen}, {Basri},
  {Charbonneau}, {Christiansen}, {Clarke}, {Dupree}, {Fabrycky}, {Fischer},
  {Ford}, {Fortney}, {Tarter}, {Girouard}, {Holman}, {Johnson}, {Klaus},
  {Machalek}, {Moorhead}, {Morehead}, {Ragozzine}, {Tenenbaum}, {Twicken},
  {Quinn}, {Isaacson}, {Shporer}, {Lucas}, {Walkowicz}, {Welsh}, {Boss},
  {Devore}, {Gould}, {Smith}, {Morris}, {Prsa}, {Morton}, {Still}, {Thompson},
  {Mullally}, {Endl}, \& {MacQueen}}]{Howard+2012}
{Howard}, A.~W., {Marcy}, G.~W., {Bryson}, S.~T., {et~al.} 2012, \apjs, 201,
  15, \dodoi{10.1088/0067-0049/201/2/15}

\bibitem[{{Hwang} {et~al.}(2018){Hwang}, {Udalski}, {Shvartzvald}, {Ryu},
  {Albrow}, {Chung}, {Gould}, {Han}, {Jung}, {Shin}, {Yee}, {Zhu}, {Cha},
  {Kim}, {Kim}, {Kim}, {Lee}, {Lee}, {Lee}, {Park}, {Pogge}, {KMTNet
  Collaboration}, {Skowron}, {Mr{\'o}z}, {Poleski}, {Koz{\l}owski},
  {Soszy{\'n}ski}, {Pietrukowicz}, {Szyma{\'n}ski}, {Ulaczyk}, {Pawlak}, {OGLE
  Collaboration}, {Bryden}, {Beichman}, {Calchi Novati}, {Gaudi}, {Henderson},
  {Jacklin}, {Penny}, \& {UKIRT Microlensing Team}}]{Hwang+2018}
{Hwang}, K.~H., {Udalski}, A., {Shvartzvald}, Y., {et~al.} 2018, \aj, 155, 20,
  \dodoi{10.3847/1538-3881/aa992f}

\bibitem[{{Ida} \& {Lin}(2004)}]{Ida+2004}
{Ida}, S., \& {Lin}, D.~N.~C. 2004, \apj, 604, 388, \dodoi{10.1086/381724}

\bibitem[{{Johnson} {et~al.}(2020){Johnson}, {Penny}, {Gaudi}, {Kerins},
  {Rattenbury}, {Robin}, {Calchi Novati}, \& {Henderson}}]{Johnson+2020}
{Johnson}, S.~A., {Penny}, M., {Gaudi}, B.~S., {et~al.} 2020, \aj, 160, 123,
  \dodoi{10.3847/1538-3881/aba75b}

\bibitem[{{Kenyon} \& {Hartmann}(1995)}]{Kenyon+1995}
{Kenyon}, S.~J., \& {Hartmann}, L. 1995, \apjs, 101, 117,
  \dodoi{10.1086/192235}

\bibitem[{{Koshimoto} {et~al.}(2021{\natexlab{a}}){Koshimoto}, {Baba}, \&
  {Bennett}}]{Koshimoto+2021}
{Koshimoto}, N., {Baba}, J., \& {Bennett}, D.~P. 2021{\natexlab{a}}, arXiv
  e-prints, arXiv:2104.03306.
\newblock \doarXiv{2104.03306}

\bibitem[{{Koshimoto} {et~al.}(2021{\natexlab{b}}){Koshimoto}, {Bennett},
  {Suzuki}, \& {Bond}}]{Koshimoto+2021b}
{Koshimoto}, N., {Bennett}, D.~P., {Suzuki}, D., \& {Bond}, I.~A.
  2021{\natexlab{b}}, \apjl, 918, L8, \dodoi{10.3847/2041-8213/ac17ec}

\bibitem[{{Koshimoto} \& {Ranc}(2021)}]{Koshimoto&Ranc2021}
{Koshimoto}, N., \& {Ranc}, C. 2021, {nkoshimoto/genulens: A Tool for
  Gravitational Microlensing Events Simulation}, v1.0,  Zenodo,
  \dodoi{10.5281/zenodo.4784949}

\bibitem[{{Koshimoto} {et~al.}(2017){Koshimoto}, {Udalski}, {Beaulieu}, {Sumi},
  {Bennett}, {Bond}, {Rattenbury}, {Fukui}, {Batista}, {Marquette}, {Brillant},
  {and}, {Abe}, {Asakura}, {Bhattacharya}, {Donachie}, {Freeman}, {Hirao},
  {Itow}, {Li}, {Ling}, {Masuda}, {Matsubara}, {Matsuo}, {Muraki}, {Nagakane},
  {Ohnishi}, {Oyokawa}, {Saito}, {Sharan}, {Shibai}, {Sullivan}, {Suzuki},
  {Tristram}, {Yonehara}, {MOA Collaboration}, {Koz{\l}owski}, {Pietrukowicz},
  {Poleski}, {Skowron}, {Soszy{\'n}ski}, {Szyma{\'n}ski}, {Ulaczyk},
  {Wyrzykowski}, \& {OGLE Collaboration}}]{Koshimoto+2017}
{Koshimoto}, N., {Udalski}, A., {Beaulieu}, J.~P., {et~al.} 2017, \aj, 153, 1,
  \dodoi{10.3847/1538-3881/153/1/1}

\bibitem[{{Kroupa} {et~al.}(1993){Kroupa}, {Tout}, \& {Gilmore}}]{Kroupa+1993}
{Kroupa}, P., {Tout}, C.~A., \& {Gilmore}, G. 1993, \mnras, 262, 545,
  \dodoi{10.1093/mnras/262.3.545}

\bibitem[{{Lingam} {et~al.}(2020){Lingam}, {Ginsburg}, \& {Loeb}}]{Lingam+2020}
{Lingam}, M., {Ginsburg}, I., \& {Loeb}, A. 2020, \apj, 888, 102,
  \dodoi{10.3847/1538-4357/ab5b13}

\bibitem[{{Lissauer}(1993)}]{Lissauer1993}
{Lissauer}, J.~J. 1993, \araa, 31, 129,
  \dodoi{10.1146/annurev.aa.31.090193.001021}

\bibitem[{{Mao} \& {Paczynski}(1991)}]{Mao+1991}
{Mao}, S., \& {Paczynski}, B. 1991, \apjl, 374, L37, \dodoi{10.1086/186066}

\bibitem[{{Marcy} {et~al.}(2005){Marcy}, {Butler}, {Fischer}, {Vogt}, {Wright},
  {Tinney}, \& {Jones}}]{Marcy+2005}
{Marcy}, G., {Butler}, R.~P., {Fischer}, D., {et~al.} 2005, Progress of
  Theoretical Physics Supplement, 158, 24, \dodoi{10.1143/PTPS.158.24}

\bibitem[{{Miyazaki} {et~al.}(2021){Miyazaki}, {Johnson}, {Sumi}, {Penny},
  {Koshimoto}, \& {Yamawaki}}]{Miyazaki+2021}
{Miyazaki}, S., {Johnson}, S.~A., {Sumi}, T., {et~al.} 2021, \aj, 161, 84,
  \dodoi{10.3847/1538-3881/abcec2}

\bibitem[{{Miyazaki} {et~al.}(2018){Miyazaki}, {Sumi}, {Bennett}, {Gould},
  {Udalski}, {Bond}, {Koshimoto}, {Nagakane}, {Rattenbury}, {Abe},
  {Bhattacharya}, {Barry}, {Donachie}, {Fukui}, {Hirao}, {Itow}, {Kawasaki},
  {Li}, {Ling}, {Matsubara}, {Matsuo}, {Muraki}, {Ohnishi}, {Ranc}, {Saito},
  {Sharan}, {Shibai}, {Suematsu}, {Suzuki}, {Sullivan}, {Tristram}, {Yamada},
  {Yonehara}, {MOA Collaboration}, {Koz{\L}owski}, {Mr{\'o}z}, {Pawlak},
  {Poleski}, {Pietrukowicz}, {Skowron}, {Soszy{\'n}ski}, {Szyma{\'n}ski},
  {Ulaczyk}, {OGLE Collaboration}, {Albrow}, {Chung}, {Han}, {Jung}, {Hwang},
  {Ryu}, {Shin}, {Shvartzvald}, {Yee}, {Zang}, {Zhu}, {Cha}, {Kim}, {Kim},
  {Kim}, {Lee}, {Lee}, {Lee}, {Park}, {Pogge}, \& {KMTNet
  Collaboration}}]{Miyazaki+2018}
{Miyazaki}, S., {Sumi}, T., {Bennett}, D.~P., {et~al.} 2018, \aj, 156, 136,
  \dodoi{10.3847/1538-3881/aad5ee}

\bibitem[{{Miyazaki} {et~al.}(2020){Miyazaki}, {Sumi}, {Bennett}, {Udalski},
  {Shvartzvald}, {Street}, {Bozza}, {Yee}, {Bond}, {Rattenbury}, {Koshimoto},
  {Suzuki}, {Fukui}, {Abe}, {Bhattacharya}, {Barry}, {Donachie}, {Fujii},
  {Hirao}, {Itow}, {Kamei}, {Kondo}, {Li}, {Ling}, {Matsubara}, {Matsuo},
  {Muraki}, {Nagakane}, {Ohnishi}, {Ranc}, {Saito}, {Sharan}, {Shibai},
  {Suematsu}, {Sullivan}, {Tristram}, {Yamakawa}, {Yonehara}, {MOA
  Collaboration}, {Skowron}, {Poleski}, {Mr{\'o}z}, {Szyma{\'n}ski},
  {Soszy{\'n}ski}, {Pietrukowicz}, {Koz{\L}owski}, {Ulaczyk}, {Wyrzykowski},
  {OGLE Collaboration}, {Friedmann}, {Kaspi}, {Maoz}, {Wise Team}, {Albrow},
  {Christie}, {DePoy}, {Gal-Yam}, {Gould}, {Lee}, {Manulis}, {McCormick},
  {Natusch}, {Ngan}, {Pogge}, {Porritt}, {{\ensuremath{\mu}}FUN Collaboration},
  {Tsapras}, {Bachelet}, {Hundertmark}, {Dominik}, {Bramich}, {Cassan},
  {Jaimes}, {Horne}, {Schmidt}, {Snodgrass}, {Wambsganss}, {Steele}, {Menzies},
  {Mao}, {RoboNet Collabofratino}, {J{\o}rgensen}, {Burgdorf}, {Ciceri},
  {Novati}, {D'Ago}, {Evans}, {Hinse}, {Kains}, {Kerins}, {Korhonen},
  {Mancini}, {Popovas}, {Rabus}, {Rahvar}, {Scarpetta}, {Skottfelt},
  {Southworth}, {D'Ago}, {Peixinho}, {Verma}, \& {MiNDSTEp
  Collaboration}}]{Miyazaki+2020}
---. 2020, \aj, 159, 76, \dodoi{10.3847/1538-3881/ab64de}

\bibitem[{{Mr{\'o}z} {et~al.}(2017){Mr{\'o}z}, {Udalski}, {Skowron}, {Poleski},
  {Koz{\l}owski}, {Szyma{\'n}ski}, {Soszy{\'n}ski}, {Wyrzykowski},
  {Pietrukowicz}, {Ulaczyk}, {Skowron}, \& {Pawlak}}]{Mroz+2017}
{Mr{\'o}z}, P., {Udalski}, A., {Skowron}, J., {et~al.} 2017, \nat, 548, 183,
  \dodoi{10.1038/nature23276}

\bibitem[{{Muraki} {et~al.}(2011){Muraki}, {Han}, {Bennett}, {Suzuki},
  {Monard}, {Street}, {Jorgensen}, {Kundurthy}, {Skowron}, {Becker}, {Albrow},
  {Fouqu{\'e}}, {Heyrovsk{\'y}}, {Barry}, {Beaulieu}, {Wellnitz}, {Bond},
  {Sumi}, {Dong}, {Gaudi}, {Bramich}, {Dominik}, {Abe}, {Botzler}, {Freeman},
  {Fukui}, {Furusawa}, {Hayashi}, {Hearnshaw}, {Hosaka}, {Itow}, {Kamiya},
  {Korpela}, {Kilmartin}, {Lin}, {Ling}, {Makita}, {Masuda}, {Matsubara},
  {Miyake}, {Nishimoto}, {Ohnishi}, {Perrott}, {Rattenbury}, {Saito},
  {Skuljan}, {Sullivan}, {Sweatman}, {Tristram}, {Wada}, {Yock}, {MOA
  Collaboration}, {Christie}, {DePoy}, {Gorbikov}, {Gould}, {Kaspi}, {Lee},
  {Mallia}, {Maoz}, {McCormick}, {Moorhouse}, {Natusch}, {Park}, {Pogge},
  {Polishook}, {Shporer}, {Thornley}, {Yee}, {{\ensuremath{\mu}}FUN
  Collaboration}, {Allan}, {Browne}, {Horne}, {Kains}, {Snodgrass}, {Steele},
  {Tsapras}, {RoboNet Collaboration}, {Batista}, {Bennett}, {Brillant},
  {Caldwell}, {Cassan}, {Cole}, {Corrales}, {Coutures}, {Dieters}, {Dominis
  Prester}, {Donatowicz}, {Greenhill}, {Kubas}, {Marquette}, {Martin},
  {Menzies}, {Sahu}, {Waldman}, {Williams}, {Zub}, {PLANET Collaboration},
  {Bourhrous}, {Matsuoka}, {Nagayama}, {Oi}, {Randriamanakoto}, {IRSF
  Observers}, {Bozza}, {Burgdorf}, {Calchi Novati}, {Dreizler}, {Finet},
  {Glitrup}, {Harps{\o}e}, {Hinse}, {Hundertmark}, {Liebig}, {Maier},
  {Mancini}, {Mathiasen}, {Rahvar}, {Ricci}, {Scarpetta}, {Skottfelt},
  {Surdej}, {Southworth}, {Wambsganss}, {Zimmer}, {MiNDSTEp Consortium},
  {Udalski}, {Poleski}, {Wyrzykowski}, {Ulaczyk}, {Szyma{\'n}ski}, {Kubiak},
  {Pietrzy{\'n}ski}, {Soszy{\'n}ski}, \& {OGLE Collaboration}}]{Muraki+2011}
{Muraki}, Y., {Han}, C., {Bennett}, D.~P., {et~al.} 2011, \apj, 741, 22,
  \dodoi{10.1088/0004-637X/741/1/22}

\bibitem[{{Nataf} {et~al.}(2013){Nataf}, {Gould}, {Fouqu{\'e}}, {Gonzalez},
  {Johnson}, {Skowron}, {Udalski}, {Szyma{\'n}ski}, {Kubiak},
  {Pietrzy{\'n}ski}, {Soszy{\'n}ski}, {Ulaczyk}, {Wyrzykowski}, \&
  {Poleski}}]{Nataf+2013}
{Nataf}, D.~M., {Gould}, A., {Fouqu{\'e}}, P., {et~al.} 2013, \apj, 769, 88,
  \dodoi{10.1088/0004-637X/769/2/88}

\bibitem[{{Nishiyama} {et~al.}(2009){Nishiyama}, {Tamura}, {Hatano}, {Kato},
  {Tanab{\'e}}, {Sugitani}, \& {Nagata}}]{Nishiyama+2009}
{Nishiyama}, S., {Tamura}, M., {Hatano}, H., {et~al.} 2009, \apj, 696, 1407,
  \dodoi{10.1088/0004-637X/696/2/1407}

\bibitem[{{Paczynski}(1986)}]{Paczynski1986}
{Paczynski}, B. 1986, \apj, 304, 1, \dodoi{10.1086/164140}

\bibitem[{{Penny} {et~al.}(2019){Penny}, {Gaudi}, {Kerins}, {Rattenbury},
  {Mao}, {Robin}, \& {Calchi Novati}}]{Penny+2019}
{Penny}, M.~T., {Gaudi}, B.~S., {Kerins}, E., {et~al.} 2019, \apjs, 241, 3,
  \dodoi{10.3847/1538-4365/aafb69}

\bibitem[{{Poindexter} {et~al.}(2005){Poindexter}, {Afonso}, {Bennett},
  {Glicenstein}, {Gould}, {Szyma{\'n}ski}, \& {Udalski}}]{Poindexter+2005}
{Poindexter}, S., {Afonso}, C., {Bennett}, D.~P., {et~al.} 2005, \apj, 633,
  914, \dodoi{10.1086/468182}

\bibitem[{{Pollack} {et~al.}(1996){Pollack}, {Hubickyj}, {Bodenheimer},
  {Lissauer}, {Podolak}, \& {Greenzweig}}]{Pollack+1996}
{Pollack}, J.~B., {Hubickyj}, O., {Bodenheimer}, P., {et~al.} 1996, \icarus,
  124, 62, \dodoi{10.1006/icar.1996.0190}

\bibitem[{{Ranc} {et~al.}(2019){Ranc}, {Bennett}, {Hirao}, {Udalski}, {Han},
  {Bond}, {Yee}, {and}, {Albrow}, {Chung}, {Gould}, {Hwang}, {Jung}, {Ryu},
  {Shin}, {Shvartzvald}, {Zang}, {Zhu}, {Cha}, {Kim}, {Kim}, {Kim}, {Lee},
  {Lee}, {Lee}, {Park}, {Pogge}, {KMTNet Collaboration}, {Abe}, {Barry},
  {Bhattacharya}, {Donachie}, {Fukui}, {Itow}, {Kawasaki}, {Kondo},
  {Koshimoto}, {Li}, {Matsubara}, {Miyazaki}, {Muraki}, {Nagakane},
  {Rattenbury}, {Suematsu}, {Sullivan}, {Sumi}, {Suzuki}, {Tristram},
  {Yonehara}, {MOA Collaboration}, {Poleski}, {Mr{\'o}z}, {Skowron},
  {Szyma{\'n}ski}, {Soszy{\'n}ski}, {Koz{\l}owski}, {Pietrukowicz}, {Ulaczyk},
  \& {OGLE Collaboration}}]{Ranc+2019}
{Ranc}, C., {Bennett}, D.~P., {Hirao}, Y., {et~al.} 2019, \aj, 157, 232,
  \dodoi{10.3847/1538-3881/ab141b}

\bibitem[{{Rattenbury} {et~al.}(2002){Rattenbury}, {Bond}, {Skuljan}, \&
  {Yock}}]{Rattenbury+2002}
{Rattenbury}, N.~J., {Bond}, I.~A., {Skuljan}, J., \& {Yock}, P.~C.~M. 2002,
  \mnras, 335, 159, \dodoi{10.1046/j.1365-8711.2002.05607.x}

\bibitem[{{Sajadian} {et~al.}(2021){Sajadian}, {Rahvar}, \&
  {Kazemian}}]{Sajadian+2021}
{Sajadian}, S., {Rahvar}, S., \& {Kazemian}, F. 2021, arXiv e-prints,
  arXiv:2103.10593.
\newblock \doarXiv{2103.10593}

\bibitem[{{Sako} {et~al.}(2008){Sako}, {Sekiguchi}, {Sasaki}, {Okajima}, {Abe},
  {Bond}, {Hearnshaw}, {Itow}, {Kamiya}, {Kilmartin}, {Masuda}, {Matsubara},
  {Muraki}, {Rattenbury}, {Sullivan}, {Sumi}, {Tristram}, {Yanagisawa}, \&
  {Yock}}]{Sako+2008}
{Sako}, T., {Sekiguchi}, T., {Sasaki}, M., {et~al.} 2008, Experimental
  Astronomy, 22, 51, \dodoi{10.1007/s10686-007-9082-5}

\bibitem[{{Shin} {et~al.}(2019){Shin}, {Yee}, {Gould}, {Penny}, {Bond},
  {Albrow}, {Chung}, {Han}, {Hwang}, {Jung}, {Ryu}, {Shvartzvald}, {Cha},
  {Kim}, {Kim}, {Kim}, {Lee}, {Lee}, {Lee}, {Park}, {Pogge}, {(KMTNet
  Collaboration}, {Abe}, {Barry}, {Bennett}, {Bhattacharya}, {Donachie},
  {Fujii}, {Fukui}, {Hirao}, {Itow}, {Kamei}, {Kondo}, {Koshimoto}, {Li},
  {Matsubara}, {Miyazaki}, {Muraki}, {Nagakane}, {Ranc}, {Rattenbury},
  {Suematsu}, {Sullivan}, {Sumi}, {Suzuki}, {Tristram}, {Yamakawa}, {Yonehara},
  {(MOA Collaboration}, {Fouqu{\'e}}, {Zang}, \& {(CFHT-K2C9 Microlensing
  Collaboration}}]{Shin+2019}
{Shin}, I.~G., {Yee}, J.~C., {Gould}, A., {et~al.} 2019, \aj, 158, 199,
  \dodoi{10.3847/1538-3881/ab46a5}

\bibitem[{{Shvartzvald} {et~al.}(2016){Shvartzvald}, {Maoz}, {Udalski}, {Sumi},
  {Friedmann}, {Kaspi}, {Poleski}, {Szyma{\'n}ski}, {Skowron}, {Koz{\l}owski},
  {Wyrzykowski}, {Mr{\'o}z}, {Pietrukowicz}, {Pietrzy{\'n}ski},
  {Soszy{\'n}ski}, {Ulaczyk}, {Abe}, {Barry}, {Bennett}, {Bhattacharya},
  {Bond}, {Freeman}, {Inayama}, {Itow}, {Koshimoto}, {Ling}, {Masuda}, {Fukui},
  {Matsubara}, {Muraki}, {Ohnishi}, {Rattenbury}, {Saito}, {Sullivan},
  {Suzuki}, {Tristram}, {Wakiyama}, \& {Yonehara}}]{Shvartzvald+2016}
{Shvartzvald}, Y., {Maoz}, D., {Udalski}, A., {et~al.} 2016, \mnras, 457, 4089,
  \dodoi{10.1093/mnras/stw191}

\bibitem[{{Skowron} {et~al.}(2011){Skowron}, {Udalski}, {Gould}, {Dong},
  {Monard}, {Han}, {Nelson}, {McCormick}, {Moorhouse}, {Thornley}, {Maury},
  {Bramich}, {Greenhill}, {Koz{\l}owski}, {Bond}, {Poleski}, {Wyrzykowski},
  {Ulaczyk}, {Kubiak}, {Szyma{\'n}ski}, {Pietrzy{\'n}ski}, {Soszy{\'n}ski},
  {OGLE Collaboration}, {Gaudi}, {Yee}, {Hung}, {Pogge}, {DePoy}, {Lee},
  {Park}, {Allen}, {Mallia}, {Drummond}, {Bolt}, {{\ensuremath{\mu}}FUN
  Collaboration}, {Allan}, {Browne}, {Clay}, {Dominik}, {Fraser}, {Horne},
  {Kains}, {Mottram}, {Snodgrass}, {Steele}, {Street}, {Tsapras}, {RoboNet
  Collaboration}, {Abe}, {Bennett}, {Botzler}, {Douchin}, {Freeman}, {Fukui},
  {Furusawa}, {Hayashi}, {Hearnshaw}, {Hosaka}, {Itow}, {Kamiya}, {Kilmartin},
  {Korpela}, {Lin}, {Ling}, {Makita}, {Masuda}, {Matsubara}, {Muraki},
  {Nagayama}, {Miyake}, {Nishimoto}, {Ohnishi}, {Perrott}, {Rattenbury},
  {Saito}, {Skuljan}, {Sullivan}, {Sumi}, {Suzuki}, {Sweatman}, {Tristram},
  {Wada}, {Yock}, {MOA Collaboration}, {Beaulieu}, {Fouqu{\'e}}, {Albrow},
  {Batista}, {Brillant}, {Caldwell}, {Cassan}, {Cole}, {Cook}, {Coutures},
  {Dieters}, {Dominis Prester}, {Donatowicz}, {Kane}, {Kubas}, {Marquette},
  {Martin}, {Menzies}, {Sahu}, {Wambsganss}, {Williams}, {Zub}, \& {PLANET
  Collaboration}}]{Skowron+2011}
{Skowron}, J., {Udalski}, A., {Gould}, A., {et~al.} 2011, \apj, 738, 87,
  \dodoi{10.1088/0004-637X/738/1/87}

\bibitem[{{Skowron} {et~al.}(2018){Skowron}, {Ryu}, {Hwang}, {Udalski},
  {Mr{\'o}z}, {Koz{\l}owski}, {Soszy{\'n}ski}, {Pietrukowicz}, {Szyma{\'n}ski},
  {Poleski}, {Ulaczyk}, {Pawlak}, {Rybicki}, {Iwanek}, {Albrow}, {Chung},
  {Gould}, {Han}, {Jung}, {Shin}, {Shvartzvald}, {Yee}, {Zang}, {Zhu}, {Cha},
  {Kim}, {Kim}, {Kim}, {Lee}, {Lee}, {Lee}, {Park}, \& {Pogge}}]{Skowron+2018}
{Skowron}, J., {Ryu}, Y.~H., {Hwang}, K.~H., {et~al.} 2018, \actaa, 68, 43,
  \dodoi{10.32023/0001-5237/68.1.2}

\bibitem[{{Sumi} {et~al.}(2003){Sumi}, {Abe}, {Bond}, {Dodd}, {Hearnshaw},
  {Honda}, {Honma}, {Kan-ya}, {Kilmartin}, {Masuda}, {Matsubara}, {Muraki},
  {Nakamura}, {Nishi}, {Noda}, {Ohnishi}, {Petterson}, {Rattenbury}, {Reid},
  {Saito}, {Saito}, {Sato}, {Sekiguchi}, {Skuljan}, {Sullivan}, {Takeuti},
  {Tristram}, {Wilkinson}, {Yanagisawa}, \& {Yock}}]{Sumi+2003}
{Sumi}, T., {Abe}, F., {Bond}, I.~A., {et~al.} 2003, \apj, 591, 204,
  \dodoi{10.1086/375212}

\bibitem[{{Sumi} {et~al.}(2011){Sumi}, {Kamiya}, {Bennett}, {Bond}, {Abe},
  {Botzler}, {Fukui}, {Furusawa}, {Hearnshaw}, {Itow}, {Kilmartin}, {Korpela},
  {Lin}, {Ling}, {Masuda}, {Matsubara}, {Miyake}, {Motomura}, {Muraki},
  {Nagaya}, {Nakamura}, {Ohnishi}, {Okumura}, {Perrott}, {Rattenbury}, {Saito},
  {Sako}, {Sullivan}, {Sweatman}, {Tristram}, {Udalski}, {Szyma{\'n}ski},
  {Kubiak}, {Pietrzy{\'n}ski}, {Poleski}, {Soszy{\'n}ski}, {Wyrzykowski},
  {Ulaczyk}, \& {Microlensing Observations in Astrophysics (MOA)
  Collaboration}}]{Sumi+2011}
{Sumi}, T., {Kamiya}, K., {Bennett}, D.~P., {et~al.} 2011, \nat, 473, 349,
  \dodoi{10.1038/nature10092}

\bibitem[{{Suzuki} {et~al.}(2016){Suzuki}, {Bennett}, {Sumi}, {Bond}, {Rogers},
  {Abe}, {Asakura}, {Bhattacharya}, {Donachie}, {Freeman}, {Fukui}, {Hirao},
  {Itow}, {Koshimoto}, {Li}, {Ling}, {Masuda}, {Matsubara}, {Muraki},
  {Nagakane}, {Onishi}, {Oyokawa}, {Rattenbury}, {Saito}, {Sharan}, {Shibai},
  {Sullivan}, {Tristram}, {Yonehara}, \& {MOA Collaboration}}]{Suzuki+2016}
{Suzuki}, D., {Bennett}, D.~P., {Sumi}, T., {et~al.} 2016, \apj, 833, 145,
  \dodoi{10.3847/1538-4357/833/2/145}

\bibitem[{{Suzuki} {et~al.}(2018){Suzuki}, {Bennett}, {Ida}, {Mordasini},
  {Bhattacharya}, {Bond}, {Donachie}, {Fukui}, {Hirao}, {Koshimoto},
  {Miyazaki}, {Nagakane}, {Ranc}, {Rattenbury}, {Sumi}, {Alibert}, \&
  {Lin}}]{Suzuki+2018}
{Suzuki}, D., {Bennett}, D.~P., {Ida}, S., {et~al.} 2018, \apjl, 869, L34,
  \dodoi{10.3847/2041-8213/aaf577}

\bibitem[{{Szyma{\'n}ski} {et~al.}(2011){Szyma{\'n}ski}, {Udalski},
  {Soszy{\'n}ski}, {Kubiak}, {Pietrzy{\'n}ski}, {Poleski}, {Wyrzykowski}, \&
  {Ulaczyk}}]{Szymanski+2011}
{Szyma{\'n}ski}, M.~K., {Udalski}, A., {Soszy{\'n}ski}, I., {et~al.} 2011,
  \actaa, 61, 83.
\newblock \doarXiv{1107.4008}

\bibitem[{{Terry} {et~al.}(2021){Terry}, {Bhattacharya}, {Bennett}, {Beaulieu},
  {Koshimoto}, {Blackman}, {Bond}, {Cole}, {Henderson}, {Lu}, {Marquette},
  {Ranc}, \& {Vandorou}}]{Terry+2021}
{Terry}, S.~K., {Bhattacharya}, A., {Bennett}, D.~P., {et~al.} 2021, \aj, 161,
  54, \dodoi{10.3847/1538-3881/abcc60}

\bibitem[{{Udalski}(2003)}]{Udalski2003}
{Udalski}, A. 2003, \actaa, 53, 291.
\newblock \doarXiv{astro-ph/0401123}

\bibitem[{{Udalski} {et~al.}(2015){Udalski}, {Szyma{\'n}ski}, \&
  {Szyma{\'n}ski}}]{Udalski+2015A}
{Udalski}, A., {Szyma{\'n}ski}, M.~K., \& {Szyma{\'n}ski}, G. 2015, \actaa, 65,
  1.
\newblock \doarXiv{1504.05966}

\bibitem[{{Vandorou} {et~al.}(2020){Vandorou}, {Bennett}, {Beaulieu}, {Alard},
  {Blackman}, {Cole}, {Bhattacharya}, {Bond}, {Koshimoto}, \&
  {Marquette}}]{Vandorou+2020}
{Vandorou}, A., {Bennett}, D.~P., {Beaulieu}, J.-P., {et~al.} 2020, \aj, 160,
  121, \dodoi{10.3847/1538-3881/aba2d3}

\bibitem[{{Verde} {et~al.}(2003){Verde}, {Peiris}, {Spergel}, {Nolta},
  {Bennett}, {Halpern}, {Hinshaw}, {Jarosik}, {Kogut}, {Limon}, {Meyer},
  {Page}, {Tucker}, {Wollack}, \& {Wright}}]{Verde+2003}
{Verde}, L., {Peiris}, H.~V., {Spergel}, D.~N., {et~al.} 2003, \apjs, 148, 195,
  \dodoi{10.1086/377335}

\bibitem[{{Wozniak}(2000)}]{Wozniak2000}
{Wozniak}, P.~R. 2000, \actaa, 50, 421.
\newblock \doarXiv{astro-ph/0012143}

\bibitem[{{Yee} {et~al.}(2012){Yee}, {Shvartzvald}, {Gal-Yam}, {Bond},
  {Udalski}, {Koz{\l}owski}, {Han}, {Gould}, {Skowron}, {Suzuki}, {Abe},
  {Bennett}, {Botzler}, {Chote}, {Freeman}, {Fukui}, {Furusawa}, {Itow},
  {Kobara}, {Ling}, {Masuda}, {Matsubara}, {Miyake}, {Muraki}, {Ohmori},
  {Ohnishi}, {Rattenbury}, {Saito}, {Sullivan}, {Sumi}, {Suzuki}, {Sweatman},
  {Takino}, {Tristram}, {Wada}, {MOA Collaboration}, {Szyma{\'n}ski}, {Kubiak},
  {Pietrzy{\'n}ski}, {Soszy{\'n}ski}, {Poleski}, {Ulaczyk}, {Wyrzykowski},
  {Pietrukowicz}, {OGLE Collaboration}, {Allen}, {Almeida}, {Batista}, {Bos},
  {Christie}, {DePoy}, {Dong}, {Drummond}, {Finkelman}, {Gaudi}, {Gorbikov},
  {Henderson}, {Higgins}, {Jablonski}, {Kaspi}, {Manulis}, {Maoz}, {McCormick},
  {McGregor}, {Monard}, {Moorhouse}, {Mu{\~n}oz}, {Natusch}, {Ngan}, {Ofek},
  {Pogge}, {Santallo}, {Tan}, {Thornley}, {Shin}, {Choi}, {Park}, {Lee}, {Koo},
  \& {{\ensuremath{\mu}}FUN Collaboration}}]{Yee+2012}
{Yee}, J.~C., {Shvartzvald}, Y., {Gal-Yam}, A., {et~al.} 2012, \apj, 755, 102,
  \dodoi{10.1088/0004-637X/755/2/102}

\bibitem[{{Yee} {et~al.}(2021){Yee}, {Zang}, {Udalski}, {Ryu}, {Green},
  {Hennerley}, {Marmont}, {Sumi}, {Mao}, {Gromadzki}, {Mr{\'o}z}, {Skowron},
  {Poleski}, {Szyma{\'n}ski}, {Soszy{\'n}ski}, {Pietrukowicz}, {Koz{\l}owski},
  {Ulaczyk}, {Rybicki}, {Iwanek}, {Wrona}, {Albrow}, {Chung}, {Gould}, {Han},
  {Hwang}, {Jung}, {Kim}, {Shin}, {Shvartzvald}, {Cha}, {Kim}, {Kim}, {Lee},
  {Lee}, {Lee}, {Park}, {Pogge}, {Bachelet}, {Christie}, {Hundertmark}, {Maoz},
  {McCormick}, {Natusch}, {Penny}, {Street}, {Tsapras}, {Beichman}, {Bryden},
  {Novati}, {Carey}, {Gaudi}, {Henderson}, {Johnson}, {Zhu}, {Bond}, {Abe},
  {Barry}, {Bennett}, {Bhattacharya}, {Donachie}, {Fujii}, {Fukui}, {Hirao},
  {Silva}, {Itow}, {Kirikawa}, {Kondo}, {Koshimoto}, {Alex Li}, {Matsubara},
  {Muraki}, {Miyazaki}, {Olmschenk}, {Ranc}, {Rattenbury}, {Satoh}, {Shoji},
  {Suzuki}, {Tanaka}, {Tristram}, {Yamawaki}, {Yonehara}, \& {MOA
  Collaboration}}]{Yee+2021}
{Yee}, J.~C., {Zang}, W., {Udalski}, A., {et~al.} 2021, \aj, 162, 180,
  \dodoi{10.3847/1538-3881/ac1582}

\bibitem[{{Yoo} {et~al.}(2004){Yoo}, {DePoy}, {Gal-Yam}, {Gaudi}, {Gould},
  {Han}, {Lipkin}, {Maoz}, {Ofek}, {Park}, {Pogge}, {Mu-Fun Collaboration},
  {Udalski}, {Soszy{\'n}ski}, {Wyrzykowski}, {Kubiak}, {Szyma{\'n}ski},
  {Pietrzy{\'n}ski}, {Szewczyk}, {{\.Z}ebru{\'n}}, \& {OGLE
  Collaboration}}]{Yoo+2004}
{Yoo}, J., {DePoy}, D.~L., {Gal-Yam}, A., {et~al.} 2004, \apj, 603, 139,
  \dodoi{10.1086/381241}

\bibitem[{{Zang} {et~al.}(2020){Zang}, {Shvartzvald}, {Udalski}, {Yee}, {Lee},
  {Sumi}, {Zhang}, {Yang}, {Mao}, {Calchi Novati}, {Gould}, {Zhu}, {Beichman},
  {Bryden}, {Carey}, {Gaudi}, {Henderson}, {Mr{\'o}z}, {Skowron}, {Poleski},
  {Szyma{\'n}ski}, {Soszy{\'n}ski}, {Pietrukowicz}, {Koz{\l}owski}, {Ulaczyk},
  {Rybicki}, {Iwanek}, {Wrona}, {Albrow}, {Chung}, {Han}, {Hwang}, {Jung},
  {Ryu}, {Shin}, {Cha}, {Kim}, {Kim}, {Kim}, {Lee}, {Lee}, {Park}, {Pogge},
  {Bond}, {Abe}, {Barry}, {Bennett}, {Bhattacharya}, {Donachie}, {Fujii},
  {Fukui}, {Hirao}, {Itow}, {Kirikawa}, {Kondo}, {Koshimoto}, {Li},
  {Matsubara}, {Muraki}, {Miyazaki}, {Ranc}, {Rattenbury}, {Satoh}, {Shoji},
  {Suzuki}, {Tanaka}, {Tristram}, {Yamawaki}, {Yonehara}, {Bachelet},
  {Hundertmark}, {Figuera Jaimes}, {Maoz}, {Penny}, {Street}, \&
  {Tsapras}}]{Zang+2020}
{Zang}, W., {Shvartzvald}, Y., {Udalski}, A., {et~al.} 2020, arXiv e-prints,
  arXiv:2010.08732.
\newblock \doarXiv{2010.08732}

\bibitem[{{Zang} {et~al.}(2021{\natexlab{a}}){Zang}, {Hwang}, {Udalski},
  {Wang}, {Zhu}, {Sumi}, {Yee}, {Gould}, {Mao}, {Zhang}, {Albrow}, {Chung},
  {Han}, {Jung}, {Ryu}, {Shin}, {Shvartzvald}, {Cha}, {Kim}, {Kim}, {Kim},
  {Lee}, {Lee}, {Lee}, {Park}, {Pogge}, {Mr{\'o}z}, {Skowron}, {Poleski},
  {Szyma{\'n}ski}, {Soszy{\'n}ski}, {Pietrukowicz}, {Koz{\l}owski}, {Ulaczyk},
  {Rybicki}, {Iwanek}, {Wrona}, {Gromadzki}, {Bond}, {Abe}, {Barry}, {Bennett},
  {Bhattacharya}, {Donachie}, {Fujii}, {Fukui}, {Hirao}, {Itow}, {Kirikawa},
  {Kondo}, {Koshimoto}, {Li}, {Matsubara}, {Muraki}, {Miyazaki}, {Olmschenk},
  {Ranc}, {Rattenbury}, {Satoh}, {Shoji}, {Ishitani Silva}, {Suzuki}, {Tanaka},
  {Tristram}, {Yamawaki}, {Yonehara}, {Beichman}, {Bryden}, {Calchi Novati},
  {Carey}, {Gaudi}, {Henderson}, {Johnson}, \& {Spitzer Team}}]{Zang+2021}
{Zang}, W., {Hwang}, K.-H., {Udalski}, A., {et~al.} 2021{\natexlab{a}}, \aj,
  162, 163, \dodoi{10.3847/1538-3881/ac12d4}

\bibitem[{{Zang} {et~al.}(2021{\natexlab{b}}){Zang}, {Han}, {Kondo}, {Yee},
  {Lee}, {Gould}, {Mao}, {de Almeida}, {Shvartzvald}, {Zhang}, {Albrow},
  {Chung}, {Hwang}, {Jung}, {Ryu}, {Shin}, {Cha}, {Kim}, {Kim}, {Kim}, {Lee},
  {Lee}, {Park}, {Pogge}, {Drummond}, {Tan}, {Nascimento J{\'u}nior}, {Maoz},
  {Penny}, {Zhu}, {Bond}, {Abe}, {Barry}, {Bennett}, {Bhattacharya},
  {Donachie}, {Fujii}, {Fukui}, {Hirao}, {Itow}, {Kirikawa}, {Koshimoto}, {Alex
  Li}, {Matsubara}, {Muraki}, {Miyazaki}, {Olmschenk}, {Ranc}, {Rattenbury},
  {Satoh}, {Shoji}, {Silva}, {Sumi}, {Suzuki}, {Tanaka}, {Tristram},
  {Yamawaki}, {Yonehara}, {Petric}, {Burdullis}, \& {Fouqu{\'e}}}]{Zang+2021b}
{Zang}, W., {Han}, C., {Kondo}, I., {et~al.} 2021{\natexlab{b}}, Research in
  Astronomy and Astrophysics, 21, 239, \dodoi{10.1088/1674-4527/21/9/239}

\bibitem[{{Zhang} {et~al.}(2020){Zhang}, {Zang}, {Udalski}, {Gould}, {Ryu},
  {Wang}, {Yang}, {Mao}, {Mr{\'o}z}, {Skowron}, {Poleski}, {Szyma{\'n}ski},
  {Soszy{\'n}ski}, {Pietrukowicz}, {Koz{\l}owski}, {Ulaczyk}, {Albrow},
  {Chung}, {Han}, {Hwang}, {Jung}, {Shin}, {Shvartzvald}, {Yee}, {Zhu}, {Cha},
  {Kim}, {Kim}, {Kim}, {Lee}, {Lee}, {Lee}, {Park}, \& {Pogge}}]{Zhang+2020}
{Zhang}, X., {Zang}, W., {Udalski}, A., {et~al.} 2020, \aj, 159, 116,
  \dodoi{10.3847/1538-3881/ab6f6d}

\end{thebibliography}
\end{document}